\documentstyle{article}

\def\noheaderplainsetup{%

      \topmargin=0pt \headheight=0pt \headsep=0pt  
      \oddsidemargin=0pt \evensidemargin=0pt       
      \textheight=8.9truein \textwidth=6.5truein}  

\noheaderplainsetup

\begin{document}


\newcommand{\concl}{\mbox{\LARGE $\mapsto$}}
\newcommand{\clt}{\mbox{\bf CL2}}
\newcommand{\valid}{\mbox{$\vdash\hspace{-4pt}\vdash$}}
\newcommand{\uvalid}{\mbox{$\vdash\hspace{-5pt}\vdash\hspace{-5pt}\vdash$}}
\newcommand{\hi}{\mbox{\bf HI}}
\newcommand{\ma}{\mbox{\bf Move alphabet}}
\newcommand{\mov}{\mbox{\em Nonrep}}
\newcommand{\rep}{\mbox{\em Rep}}
\newcommand{\chess}{\mbox{\em Chess}}
\newcommand{\checkers}{\mbox{\em Checkers}}
\newcommand{\rneg}{\neg}               
\newcommand{\pneg}{\neg}               
\newcommand{\emptyrun}{\langle\rangle} 
\newcommand{\oo}{\bot}            
\newcommand{\pp}{\top}            
\newcommand{\xx}{\wp}               
\newcommand{\s}[1]{\underline{#1}} 
\newcommand{\blu}[1]{\overline{#1}} 
\newcommand{\yel}[1]{\underline{#1}} 
\newcommand{\cont}[1]{\underline{\overline{#1}}} 
\newcommand{\leg}[3]{\mbox{\bf Lm}{#2}^{#1}_{#3}} 
\newcommand{\legal}[2]{\mbox{\bf Lr}^{#1}_{#2}} 
\newcommand{\Legal}[1]{\mbox{\bf LR}^{#1}} 
\newcommand{\win}[2]{\mbox{\bf Wn}^{#1}_{#2}} 
\newcommand{\tree}[2]{\mbox{\em Tree}\seq{#2}} 
\newcommand{\constants}{\mbox{\bf Constants}} 
\newcommand{\variables}{\mbox{\bf Variables}}    
\newcommand{\moves}{\mbox{\bf Moves}}        
\newcommand{\runs}{\mbox{\bf Runs}} 
\newcommand{\players}{\mbox{\bf Players}} 
\newcommand{\instances}{\mbox{\bf Valuations}} 
\newcommand{\bpar}[1]{\mbox{$\bigl( #1 \bigr)$}}            
\newcommand{\seq}[1]{\langle #1 \rangle}           
\newcommand{\tuple}[2]{\mbox{$ {#1}_{1},\ldots,{#1}_{#2}$}} 
\newcommand{\col}[1]{\mbox{$#1$:}} 
\newcommand{\coli}[1]{\mbox{\scriptsize $#1$:}}
\newcommand{\hint}{\mbox{\bf INT}}


\newcommand{\sst}{\mbox{\raisebox{-0.07cm}{\scriptsize $-$}\hspace{-0.2cm}$\pst$}}

\newcommand{\pst}{\mbox{\raisebox{-0.01cm}{\scriptsize $\wedge$}\hspace{-4pt}\raisebox{0.16cm}{\tiny $\mid$}\hspace{2pt}}}
\newcommand{\gneg}{\neg}                  
\newcommand{\intimpl}{\mbox{\hspace{2pt}$\circ$\hspace{-0.14cm} \raisebox{-0.058cm}{\Large --}\hspace{2pt}}}
\newcommand{\mli}{\rightarrow}                     
\newcommand{\mleq}{\hspace{2pt}\leftrightarrow\hspace{2pt}}   
\newcommand{\cla}{\mbox{\large $\forall$}}      
\newcommand{\cle}{\mbox{\large $\exists$}}        
\newcommand{\mld}{\vee}    
\newcommand{\mlc}{\wedge}  
\newcommand{\ade}{\mbox{\Large $\sqcup$}}      
\newcommand{\ada}{\mbox{\Large $\sqcap$}}      
\newcommand{\add}{\sqcup}                      
\newcommand{\adc}{\sqcap}                      
\newcommand{\clai}{\forall}     
\newcommand{\clei}{\exists}        
\newcommand{\adei}{\mbox{$\sqcup$}}      
\newcommand{\adai}{\mbox{$\sqcap$}}      
\newcommand{\sti}{\mbox{\raisebox{-0.02cm}
{\scriptsize $\circ$}\hspace{-0.121cm}\raisebox{0.08cm}{\tiny $.$}\hspace{-0.079cm}\raisebox{0.10cm}
{\tiny $.$}\hspace{-0.079cm}\raisebox{0.12cm}{\tiny $.$}\hspace{-0.085cm}\raisebox{0.14cm}
{\tiny $.$}\hspace{-0.079cm}\raisebox{0.16cm}{\tiny $.$}\hspace{1pt}}}
\newcommand{\costi}{\mbox{\raisebox{0.08cm}
{\scriptsize $\circ$}\hspace{-0.121cm}\raisebox{-0.01cm}{\tiny $.$}\hspace{-0.079cm}\raisebox{0.01cm}
{\tiny $.$}\hspace{-0.079cm}\raisebox{0.03cm}{\tiny $.$}\hspace{-0.085cm}\raisebox{0.05cm}
{\tiny $.$}\hspace{-0.079cm}\raisebox{0.07cm}{\tiny $.$}\hspace{1pt}}}
\newcommand{\intf}{\$}               
\newcommand{\tlg}{\bot}               
\newcommand{\twg}{\top}               
\newcommand{\st}{\mbox{\raisebox{-0.05cm}{$\circ$}\hspace{-0.13cm}\raisebox{0.16cm}{\tiny $\mid$}\hspace{2pt}}}
\newcommand{\cost}{\mbox{\raisebox{0.12cm}{$\circ$}\hspace{-0.13cm}\raisebox{0.02cm}{\tiny $\mid$}\hspace{2pt}}}
\newcommand{\pcost}{\mbox{\raisebox{0.12cm}{\scriptsize $\vee$}\hspace{-4pt}\raisebox{0.02cm}{\tiny $\mid$}\hspace{2pt}}}
\newcommand{\ssti}
{\mbox{\raisebox{-0.04cm}
{\tiny --}\hspace{-0.24cm}
{\tiny $\wedge$}\hspace{-0.135cm}\raisebox{0.08cm}{\tiny $.$}\hspace{-0.079cm}\raisebox{0.10cm}
{\tiny $.$}\hspace{-0.079cm}\raisebox{0.12cm}{\tiny $.$}\hspace{-0.085cm}\raisebox{0.14cm}
{\tiny $.$}\hspace{-0.079cm}\raisebox{0.16cm}{\tiny $.$}\hspace{1pt}}}
\newcommand{\scosti}
{\mbox{\raisebox{0.14cm}
{\tiny --}\hspace{-0.24cm}
\raisebox{0.08cm}
{\tiny $\vee$}\hspace{-0.135cm}\raisebox{-0.01cm}{\tiny $.$}\hspace{-0.079cm}\raisebox{0.01cm}
{\tiny $.$}\hspace{-0.079cm}\raisebox{0.03cm}{\tiny $.$}\hspace{-0.085cm}\raisebox{0.05cm}
{\tiny $.$}\hspace{-0.079cm}\raisebox{0.07cm}{\tiny $.$}\hspace{1pt}}}


\newtheorem{theoremm}{Theorem}[section]
\newtheorem{thesiss}[theoremm]{Thesis}
\newtheorem{definitionn}[theoremm]{Definition}
\newtheorem{figuree}[theoremm]{Figure}
\newtheorem{lemmaa}[theoremm]{Lemma}
\newtheorem{propositionn}[theoremm]{Proposition}
\newtheorem{conventionn}[theoremm]{Convention}
\newtheorem{examplee}[theoremm]{Example}
\newtheorem{remarkk}[theoremm]{Remark}
\newtheorem{conjecturee}[theoremm]{Conjecture}
\newtheorem{claimm}[theoremm]{Claim}
\newtheorem{corollaryy}[theoremm]{Corollary}

\newenvironment{definition}{\begin{definitionn} \em}{ \end{definitionn}}
\newenvironment{theorem}{\begin{theoremm}}{\end{theoremm}}
\newenvironment{lemma}{\begin{lemmaa}}{\end{lemmaa}}
\newenvironment{corollary}{\begin{corollaryy}}{\end{corollaryy}}
\newenvironment{proposition}{\begin{propositionn} }{\end{propositionn}}
\newenvironment{convention}{\begin{conventionn} \em}{\end{conventionn}}
\newenvironment{remark}{\begin{remarkk} \em}{\end{remarkk}}
\newenvironment{proof}{ {\bf Proof.} }{\  $\Box$ \vspace{.1in} }
\newenvironment{conjecture}{\begin{conjecturee} }{\end{conjecturee}}
\newenvironment{claim}{\begin{claimm} }{\end{claimm}}
\newenvironment{thesis}{\begin{thesiss} }{\end{thesiss}}
\newenvironment{example}{\begin{examplee} \em}{\end{examplee}}
\newenvironment{figuret}{\begin{figuree} \em}{\end{figuree}}

\title{Intuitionistic computability logic}
\author{Giorgi Japaridze\thanks{This material is based upon work supported by the National Science Foundation under Grant No. 0208816} 
}
\date{}
\maketitle

\begin{abstract} Computability logic (CL) is a systematic formal theory of computational tasks and resources, which,
in a sense, can be seen as a semantics-based alternative to (the syntactically introduced) linear logic. 
With its expressive and flexible language, where formulas represent computational problems and ``truth" is understood as 
algorithmic solvability, CL potentially offers a comprehensive logical basis for constructive applied theories and
computing systems inherently requiring constructive and computationally meaningful underlying logics.  
 Among the best known 
constructivistic logics is Heyting's intuitionistic calculus $\hint$, whose language can be seen as a special fragment of that of CL. The constructivistic philosophy of $\hint$, however, just like the resource philosophy of linear logic, has never really found an intuitively convincing and mathematically strict semantical justification. CL has good claims to provide such a justification and hence a materialization of Kolmogorov's known thesis ``$\hint$ = logic of problems".  
The present paper contains a soundness proof for $\hint$ with respect to 
the CL semantics. 
\end{abstract}




\section{Introduction}\label{intr}

{\em Computability logic} (CL), introduced recently in \cite{Jap03}, is a formal theory of computability in the same sense as
classical logic is a formal theory of truth. It understands formulas as (interactive) computational problems, and their 
``truth" as algorithmic solvability. Computational problems, in turn, are defined as games played by a
machine against the environment, with algorithmic solvability meaning existence of a machine that always wins the game.  

Intuitionistic computability logic is not a modification or version of CL. The latter takes pride in its universal applicability, stability and ``immunity to possible future revisions and tampering`` (\cite{Jap03}, p. 12). Rather,  what we refer to as  {\em intuitionistic computability logic} is just a --- relatively modest --- fragment of CL, obtained by mechanically restricting its formalism to a special sublanguage. 
It was conjectured in \cite{Jap03} that the (set of the valid formulas of the) resulting fragment of CL is described by Heyting's {\em intuitionistic calculus} $\hint$. The present paper is devoted to a verification of the soundness part of that conjecture. 

Bringing $\hint$ and CL together could  signify a step forward not only in logic but also in theoretical computer science. $\hint$ has been attracting the attention of computer scientists since long ago. And not only due to the 
beautiful phenomenon within the `formulas-as-types' approach known as the Curry-Howard isomorphism. $\hint$ appears to be an appealing alternative to classical logic within 
the more traditional approaches as well.
This is due to the general constructive features of its deductive machinery, and Kolmogorov's \cite{Kol32} well-known  yet so far rather abstract thesis according to which intuitionistic logic is
(or should be) a logic of problems. The latter inspired many attempts  
to find a ``problem semantics" for the language of intuitionistic logic \cite{God58,Kle52,Med63}, none of which, however, has fully succeeded in justifying $\hint$ as a logic of problems. Finding a semantical justification 
for $\hint$ was also among the main motivations for Lorenzen \cite{Lor59}, who pioneered game-semantical approaches in logic. After a couple of decades of trial and error, the goal of obtaining soundness and completeness of $\hint$ with respect to Lorenzen's game semantics was achieved \cite{Fel85}. The value of such an achievement is, however, dubious, as it 
came as  
a result of carefully tuning the semantics and adjusting it to the goal at the cost of sacrificing some natural  
intuitions that a game semantics could potentially offer.\footnote{Using Blass's \cite{Bla92} words, `Supplementary rules governing repeated attacks and defenses were devised by Lorenzen so that the formulas for which $P$ [proponent] has a winning strategy are exactly the intuitionistically provable ones'. Quoting \cite{Jap97}, `Lorenzen's approach describes logical validity exclusively in terms of rules without appealing 
to any kind of truth values for atoms, and this makes the semantics somewhat vicious ... as it looks like just a ``pure" syntax rather than a semantics'.}
After all, some sort of a specially designed technical semantics can be found for virtually every  formal system,
but the whole question is how natural and usable such a semantics is in its own right. In contrast, 
the CL semantics was elaborated without any target deductive construction in mind, following the motto ``{\em Axiomatizations should serve meaningful semantics rather than vice versa}". Only retroactively was it observed that the semantics of CL yields logics similar to or identical with some known axiomatically introduced constructivistic logics such as linear logic or $\hint$. Discussions given in \cite{Jap03,Japtcs,Japic,Japtocl1} demonstrate how naturally the semantics 
of CL emerges and how much utility it offers, with potential application areas ranging 
from the pure theory of (interactive) computation to knowledgebase systems, systems for planning and action, and constructive applied theories. As this semantics has well-justified claims to be a semantics of computational problems, the 
results of the present article speak strongly in favor of Kolmogorov's thesis, with a promise of a full materialization of the thesis in case a completeness proof of $\hint$ is also found. 

The main utility of the present result is in the possibility to base applied theories or knowledgebase systems on $\hint$. Nonlogical axioms --- or the knowledge base --- of such a system would be any collection of (formulas expressing) problems whose algorithmic solutions are known. Then, our soundness theorem for $\hint$ --- which comes in a strong form called uniform-constructive soundness --- guarantees that every theorem $T$ of the theory also has an algorithmic solution and, furthermore, such
a solution can be effectively constructed from a proof of $T$. This makes $\hint$ a  problem-solving tool: finding a solution for a given problem reduces to finding a proof of that problem in the theory. 

It is not an ambition of the present paper to motivationally (re)introduce and (re)justify computability logic and its intuitionistic fragment in particular. This job has been done in \cite{Jap03} and once again --- in a more compact way --- in \cite{Japic}. An assumption is that the reader is familiar with at least the motivational/philosophical parts of either paper and this is why (s)he decided to read the present article. While helpful in fully understanding the import of the present results,
from the purely technical point of view such a familiarity, however, is not necessary, as this paper provides all necessary definitions. Even if so, \cite{Jap03} and/or \cite{Japic} could still help a less advanced  reader in getting a better hold of the basic technical concepts. Those papers are written in a semitutorial style,
containing ample examples, explanations and illustrations, with \cite{Japic} even including exercises.   

\section{A brief informal overview of some basic concepts}\label{ss2}

As noted, formulas of CL represent interactive computational problems. Such problems are understood as games between two players: $\pp$, called {\bf machine}, and $\oo$, called {\bf environment}. $\pp$ is a mechanical device with a fully determined, algorithmic behavior.  On the other hand, there are no restrictions on the behavior of $\oo$. A problem/game 
is considered (algorithmically) solvable/winnable iff there is a machine that wins the game no matter 
how the environment acts. 

Logical operators are understood as operations on games/problems. One of the important groups of such operations, called {\bf choice operations}, consists 
of $\adc,\add,\ada,\ade$, in our present approach corresponding to the intuitionistic operators of conjunction, disjunction, universal quantifier and existential quantifier, respectively. $A_1\adc\ldots \adc A_n$ is a game where the first legal move (``choice"), which should be one of the elements of $\{1,\ldots,n\}$, is by the environment. After such a move/choice $i$ is made, the play continues and the winner is determined according to the rules of $A_i$; if a choice is never made, $\oo$ loses. 
 $A_1\add\ldots\add A_n$ is defined in a symmetric way with the roles of $\oo$ and $\pp$ interchanged: here it is $\pp$ who makes an initial choice and who loses if such a choice is not made. With the universe of discourse being $\{1,2,3,\ldots\}$, the meanings of the ``big brothers" $\ada$ and $\ade$ of $\adc$ and $\add$ can now be explained by 
$\ada x A(x)=A(1)\adc A(2)\adc A(3)\adc \ldots$ and $\ade x A(x)=A(1)\add A(2)\add A(3)\add\ldots$. 

The remaining two operators of intuitionistic logic are the binary $\intimpl$ (``intuitionistic implication") and the $0$-ary $\intf$ (``intuitionistic absurd"), with the intuitionistic negation of $F$ simply understood as an abbreviation for $F\intimpl \intf$. The intuitive meanings of $\intimpl$ and $\intf$ are ``reduction" (in the weakest possible sense) and ``a problem of universal strength", respectively. In what precise sense is $\intf$ a universal-strength problem  
will be seen in Section \ref{s2}. As for $\intimpl$, its meaning can be better explained in terms of some other, more basic, operations of CL that have no official intuitionistic counterparts.

One group of such operations comprises {\bf negation} $\gneg$ and the so called {\bf parallel operations} $\mlc,\mld,\mli$.
Applying $\gneg$ to a game $A$ interchanges the roles of the two players: $\pp$'s moves and wins become $\oo$'s moves and wins, and vice versa. Say, if $\chess$ is the game of chess from the point of view of the white player, then $\gneg\chess$ is the same game as seen by the black player. Playing $A_1\mlc \ldots\mlc A_n$ (resp. $A_1\mld\ldots\mld A_n$) means playing the $n$ games in parallel where, in order to win, $\pp$ needs to win in all (resp. at least one) of the components $A_i$. Back to our chess example, the two-board game $\chess\mld\gneg\chess$ can be easily won by just mimicking in $\chess$ the moves made by the adversary in $\gneg \chess$ and vice versa. On the other hand, winning $\chess \add\gneg \chess$ is not easy at all: here $\pp$ needs to choose between $\chess$ and $\gneg\chess$ (i.e. between playing white or black), and then win the chosen one-board game. Technically, a move $\alpha$ in the $k$th $\mlc$-conjunct or $\mld$-disjunct is made by prefixing $\alpha$ with `$k.$'. For example, in (the initial position of) $(A\add B)\mld (C\adc D)$, the move `$2.1$' is legal for $\oo$, meaning choosing the first $\adc$-conjunct in the 
second $\mld$-disjunct of the game. If such a move is made, the game will continue as $(A\add B)\mld C$. One of the distinguishing features of CL games from the more traditional concepts of games (\cite{Bla72,Bla92,Fel85,Jap97,Lor59}) is the absence of {\em procedural rules} --- rules strictly regulating which of the players can or should move in any given situation. E.g., in  the above game
  $(A\add B)\mld (C\adc D)$, $\pp$ also has legal moves --- the moves `$1.1$' and `$1.2$'. In such cases CL allows either player to move, depending on who wants or can act faster.\footnote{This is true for the case when the underlying 
model of computation is HPM (see Section \ref{icp}), but seemingly not so when it is EPM --- the model employed in the present paper. It should be remembered, however, that EPM is viewed as a secondary model in CL, admitted only due to the fact that it has been proven (\cite{Jap03}) to be equivalent to the basic HPM model.}  As argued in \cite{Jap03} (Section 3), only this 
``free" approach makes it possible to adequately capture certain natural intuitions such as truly parallel/concurrent computations. 

The operation $\mli$ is defined by $A\mli B=(\gneg A)\mld B$. Intuitively, this is the problem of {\em reducing} 
$B$ to $A$: solving $A\mli B$ means solving $B$ having $A$ as an external {\em computational resource}. Resources are symmetric to problems: what is a problem to solve for one player is a resource that the other player can use, and vice versa. Since 
$A$ is negated in  $(\gneg A)\mld B$ and negation means switching the roles, $A$ appears as a resource rather than problem for 
$\pp$ in $A\mli B$. To get a feel of $\mli$ as a problem reduction operation, the following --- already ``classical" in CL --- example may help. Let, for any $m,n$, {\em Accepts}$(m,n)$ mean the game where none of the players has legal moves, and which is automatically won by $\pp$ if Turing machine $m$ accepts input $n$, and otherwise automatically lost. 
This sort of zero-length games are called {\bf elementary} in CL, which understands every classical proposition/predicate
as an elementary game and vice versa, with ``true"=``won by $\pp$" and ``false"=``lost by $\pp$".  Note that then 
$\ada x\ada y\bigl(\mbox{\em Accepts}(x,y)\add\gneg \mbox{\em Accepts}(x,y)\bigr)$ expresses the acceptance problem 
as a decision problem: in order to win, the machine should be able to  tell whether $x$ accepts $y$ or not (i.e., choose the true disjunct) for any particular values 
for $x$ and $y$ selected by the environment. This problem is undecidable, which obviously means that there is no machine that (always) wins the game $\ada x\ada y\bigl(\mbox{\em Accepts}(x,y)\add\gneg \mbox{\em Accepts}(x,y)\bigr)$. However, the acceptance problem is known to be algorithmically reducible to the halting problem. The latter can be expressed by 
$\ada x\ada y\bigl(\mbox{\em Halts}(x,y)\add\gneg \mbox{\em Halts}(x,y)\bigr)$, with the obvious meaning of the elementary game/predicate  
$\mbox{\em Halts}(x,y)$. This reducibility translates into our terms as existence of a machine that wins
\begin{equation}\label{first}\ada x\ada y\bigl(\mbox{\em Halts}(x,y)\add\gneg \mbox{\em Halts}(x,y)\bigr)\mli \ada x\ada y\bigl(\mbox{\em Accepts}(x,y)\add\gneg \mbox{\em Accepts}(x,y)\bigr).\end{equation}
Such a machine indeed exists. A successful strategy for it is as follows. At the beginning, $\pp$ waits  till 
$\oo$ specifies some values $m$ and $n$ for $x$ and $y$ in the consequent, i.e. makes the moves `$2.m$' and `$2.n$'. Such moves, bringing the consequent down to $\mbox{\em Accepts}(m,n)\add\gneg \mbox{\em Accepts}(m,n)$, 
 can be seen as asking the question ``does machine $m$ accept input $n$?". To this question $\pp$ replies by the counterquestion ``does $m$ halt on $n$?", i.e. makes the moves `$1.m$ and `$1.n$', bringing the antecedent down to 
$\mbox{\em Halts}(m,n)\add\gneg \mbox{\em Halts}(m,n)$. The environment
has to correctly answer this counterquestion, or else it loses. If it answers ``no" (i.e. makes the move `$1.2$' and thus further brings the antecedent down to $\gneg \mbox{\em Halts}(m,n)$), $\pp$ also answers ``no" to the 
original question in the consequent (i.e. makes the move `$2.2$'), with the overall game having evolved to the true
and hence $\pp$-won proposition/elementary game $\gneg \mbox{\em Halts}(m,n)\mli \gneg \mbox{\em Accepts}(m,n)$. Otherwise, if the environment's answer is ``yes" 
(move `$1.1$'), $\pp$ simulates Turing machine $m$ on input $n$ until it halts, and then makes the move `$2.1$' or `$2.2$' depending whether the simulation accepted or rejected.  

Various sorts of reduction have been defined and studied in an ad hoc manner in the literature. A strong case can be made in favor of the thesis that the reduction captured by our $\mli$ is the most basic one, with all other reasonable concepts of reduction being definable in terms of $\mli$. Most natural of those concepts is the one captured by 
the earlier-mentioned operation of ``intuitionistic implication" $\intimpl$, with $A\intimpl B$ defined in terms of $\mli$ and 
(yet another natural operation) $\st$ by $A\intimpl B=(\st A)\mli B$. What makes $\intimpl$ so natural is that it captures our intuition of reducing one problem to another in the weakest possible sense. The well-established concept of Turing reduction has the same claim. But the latter is only defined for non-interactive, two-step (question/answer, or input/output) problems, such as the above halting or acceptance problems. When restricted to this sort of problems, as one might expect, $\intimpl$ indeed turns out to be equivalent to Turing reduction. The former, however, is more general than the latter as it is applicable to all problems regardless their forms and degrees of interactivity. Turing reducibility of 
a problem $B$ to a problem $A$ is defined as the possibility to algorithmically solve $B$ having an oracle for $A$. Back to (\ref{first}), 
the role of $\oo$ in the antecedent is in fact that of an oracle for the halting problem. Notice, however, that the usage of the oracle is limited there as it only can be employed once: after querying regarding whether $m$ halts of $n$, the machine would not be able to repeat the same query with different parameters $m'$ and $n'$, for that would require two 
``copies" of $\ada x\ada y\bigl(\mbox{\em Halts}(x,y)\add\gneg \mbox{\em Halts}(x,y)\bigr)$ rather than one. On the other 
hand, Turing reduction to $A$ and,  similarly, our $A\intimpl\ldots$, allow unlimited and recurring usage of $A$, which the resource-conscious CL understands as $\mli$-reduction not to $A$ but to the stronger problem expressed by  $\st A$, called the {\bf branching recurrence} of $A$.\footnote{The term ``branching recurrence" and the symbols $\sti$ and $\intimpl$ were established in \cite{Japic}. The earlier paper \cite{Jap03} uses ``branching conjunction", $!$ and $\Rightarrow$ instead. 
In the present paper, $\Rightarrow$ has a different meaning --- that of a separator of the two parts of a sequent.} 
Two more recurrence operations have been introduced within the framework of CL (\cite{Japic}): {\em parallel recurrence} $\pst$ and 
{\em sequential recurrence} $\sst$. Common to all of these operations is that, when applied to a resource $A$, they turn it into a resource that allows to reuse $A$ an unbounded number of times. The difference is in how ``reusage" is exactly understood. Imagine a computer that has a program successfully playing $\chess$. The resource that such a computer provides is obviously something stronger than just $\chess$, for it allows to play $\chess$ as many times as the user wishes, while $\chess$, as such, only assumes one play. The simplest operating system would allow to start a session of $\chess$, then --- after finishing or abandoning and destroying it --- start a new play again, and so on. The game that such a system plays --- i.e. the resource that it supports/provides --- 
is $\sst \chess$, which assumes an unbounded number of plays of $\chess$ in a sequential fashion. However, a more advanced operating system would not require to destroy the old session(s) before starting a new one; rather, it would allow to run as many parallel sessions as the user needs. This is what is captured by $\pst\chess$, meaning nothing but the infinite 
conjunction $\chess\mlc\chess\mlc\ldots$. As a resource, $\pst\chess$ is obviously stronger than $\sst\chess$ as it gives the user more flexibility. But $\pst$ is still not the strongest form of reusage. A really good operating system would not only allow the user to start new sessions of $\chess$ without destroying old ones; it would also make it possible to branch/replicate each particular session, i.e. create any number of ``copies" of any already reached position
of the multiple parallel plays of $\chess$, thus giving the user 
the possibility to try different continuations from the same position. 
After analyzing the formal definition of $\st$ given in Section \ref{cg} --- or, better, the explanations provided in Section 13 of \cite{Jap03} --- the reader will see that $\st\chess$ is exactly what accounts for this sort of a situation. $\pst\chess$ can then be thought of as a restricted 
version of $\st\chess$ where only the initial position can be replicated. A well-justified claim can be made 
that $\st A$ captures our strongest possible intuition of ``recycling"/``reusing" $A$. This automatically translates into 
another claim, according to which $A\intimpl B$, i.e. $\st A\mli B$, captures our weakest possible --- and hence most
 natural --- intuition of reducing $B$ to $A$. 

As one may expect, the three concepts of recurrence 
validate different principles. 
For example, one can show that the left $\add$- or $\ade$-introduction rules of $\hint$, which are sound with $A\intimpl B$ understood as $\st A\mli B$, would fail if $A\intimpl B$ was understood as $\pst A\mli B$ or $\sst A\mli B$. A naive person familiar with linear logic  and seeing philosophy-level connections between our recurrence operations and 
Girard's \cite{Gir87} {\em storage} operator $!$, might ask which of the three recurrence operations ``corresponds" 
to $!$. In the absence of a clear resource semantics for linear logic, perhaps such a question would not be quite meaningful though.
Closest to our present approach is that of \cite{Bla72}, where Blass proved soundness for the propositional fragment of $\hint$ with respect to his semantics, reintroduced 20 years later \cite{Bla92} in the new context of linear logic. 

To appreciate the difference between $\mli$ and $\intimpl$, let us remember the Kolmogorov complexity problem. It can be expressed by $\ada u\ade z K(z,u)$, where $K(z,u)$ is the predicate ``$z$ is the size of 
the smallest (code of a) Turing machine that returns $u$ on input $1$". Just like the acceptance problem, the Kolmogorov complexity problem has no algorithmic solution but is algorithmically reducible to the halting problem. However, such a reduction can be shown to essentially require 
recurring usage of the resource $\ada x\ada y\bigl(\mbox{\em Halts}(x,y)\add\gneg \mbox{\em Halts}(x,y)\bigr)$. That is, 
while the following game is winnable by a machine, it is not so with $\mli$ instead of $\intimpl$:\vspace{-3pt}
\begin{equation}\label{second}\ada x\ada y\bigl(\mbox{\em Halts}(x,y)\add\gneg \mbox{\em Halts}(x,y)\bigr)\intimpl
\ada u\ade z K(z,u).\vspace{-3pt}\end{equation}
Here is $\pp$'s strategy for (\ref{second}) in relaxed terms: $\pp$ waits till $\oo$ selects a value $m$ for $u$ in the consequent, thus asking $\pp$ the question ``what is the Kolmogorov complexity of $m$?". After this, starting from $i=1$, $\pp$ does the following: it creates a new copy of the (original) antecedent, and makes the two moves in it specifying $x$ and $y$ as $i$ and $1$, respectively, thus asking the counterquestion ``does machine $i$ halt on input $1$?". If $\oo$ responds by choosing $\gneg \mbox{\em Halts}(i,1)$ (``no"), $\pp$ increments $i$ by one and repeats the step; otherwise, if $\oo$ responds by $\mbox{\em Halts}(i,1)$ (``yes"), $\pp$ simulates machine $i$ on input $1$ until it halts; if it sees that machine $i$ returned $m$, it makes the move in the consequent specifying $z$ as $|i|$ (here $|i|$ means the size of $i$, i.e., $|i|= \mbox{\boldmath log}_{2} i$), thus saying that $|i|$ is the Kolmogorov complexity of $m$; otherwise, it increments $i$ by one and repeats the step. 

\section{Constant games}\label{cg}

Now we are getting down to formal definitions of the concepts informally explained in the previous section. Our ultimate concept of games will be defined in the next section in terms of the simpler and more basic class of games 
called constant games. 
To define this class, we need some technical terms and conventions. Let us agree that by a {\bf move}\label{y31} we mean any finite string over the standard keyboard alphabet. One of the non-numeric and non-punctuation symbols of the alphabet, denoted $\spadesuit$,\label{yy4} is designated as a special-status move, intuitively meaning a move that is always illegal to make. 
A {\bf labeled move}\label{x39} ({\bf labmove}) is a move prefixed with $\pp$ or $\oo$, with its prefix ({\bf label}) indicating which player has made the move. 
A {\bf run}\label{y41} is a (finite or infinite) sequence of labeled moves, and a {\bf position}\label{y40} is a finite run. 

\begin{convention}\label{conv1}
We will be exclusively using the letters $\Gamma,\Theta,\Phi,\Psi,\Upsilon$  for runs, $\xx$ for players, $\alpha,\beta,\gamma$ for moves, and $\lambda$ for labmoves. 
Runs will be often delimited with ``$\langle$" and ``$\rangle$", with $\emptyrun$\label{yy3} thus denoting the {\bf empty run}. The meaning of an expression such as $\seq{\Phi,\xx\alpha,\Gamma}$ must be clear: this is the result of appending to position $\seq{\Phi}$ 
the labmove $\seq{\xx\alpha}$ and then the run $\seq{\Gamma}$. $\rneg\Gamma$ (not to confuse this $\rneg$ with the same-shape game operation of negation) will mean the result of simultaneously replacing every label $\pp$ in every labmove of $\Gamma$ by $\oo$ and vice versa. Another important notational convention is that, for a string/move $\alpha$, 
$\Gamma^\alpha$ means  the result of removing from $\Gamma$ all labmoves 
except those of the form $\xx\alpha\beta$, and then deleting the prefix `$\alpha$' in the remaining moves, i.e. replacing each such $\xx \alpha\beta$ by $\xx\beta$. 
\end{convention}

The following item is a formal definition of constant games combined with some less formal conventions regarding the usage of certain terminology.

\begin{definition}\label{game}
 A {\bf constant game} is a pair $A=(\legal{A}{},\win{A}{})$, where:

1. $\legal{A}{}$ is a set of runs not containing (whatever-labeled) move $\spadesuit$, satisfying the condition that a (finite or infinite) run is in $\legal{A}{}$ iff all of its nonempty finite --- not necessarily proper --- initial
segments are in $\legal{A}{}$ (notice that this implies $\emptyrun\in\legal{A}{}$). The elements of $\legal{A}{}$ are
said to be {\bf legal runs} of $A$, and all other runs are said to be {\bf illegal}. We say that $\alpha$ is a {\bf legal move} for $\xx$ in a position $\Phi$ of $A$ iff $\seq{\Phi,\xx\alpha}\in\legal{A}{}$; otherwise 
$\alpha$ is {\bf illegal}. When the last move of the shortest illegal initial segment of $\Gamma$  is $\xx$-labeled, we say that $\Gamma$ is a {\bf $\xx$-illegal} run of $A$. 

2. $\win{A}{}$  is a function that sends every run $\Gamma$ to one of the players $\pp$ or $\oo$, satisfying the condition that if $\Gamma$ is a $\xx$-illegal run of $A$, then $\win{A}{}\seq{\Gamma}\not=\xx$. When $\win{A}{}\seq{\Gamma}=\xx$, we say that $\Gamma$ is a {\bf $\xx$-won} (or {\bf won by $\xx$}) run of $A$; otherwise $\Gamma$ is {\bf lost} by $\xx$. Thus, an illegal run is always lost by the player who has made the first illegal move in it.  
\end{definition}

\begin{definition}\label{op} Let $A$, $B$, $A_1,A_2,\ldots$ be constant games, and $n\in\{2,3,\ldots\}$.

1. $\gneg A$ is defined by: $\Gamma\in\legal{\gneg A}{}$ iff $\rneg{\Gamma}\in\legal{A}{}$; $\win{\gneg A}{}\seq{\Gamma}=\pp$ iff $\win{A}{}\seq{\rneg\Gamma}=\oo$.

2. $A_1\adc\ldots\adc  A_n$ is defined by: $\Gamma\in\legal{A_1\adc\ldots\adc A_n}{}$ iff $\Gamma=\emptyrun$ or $\Gamma=\seq{\oo i,\Theta}$, where $i\in\{1,\ldots,n\}$ and 
$\Theta\in\legal{A_i}{}$; $\win{A_1\adc\ldots\adc A_n}{}\seq{\Gamma}=\oo$ iff $\Gamma=\seq{\oo i,\Theta}$, where $i\in\{1,\ldots,n\}$ and $\win{A_i}{}\seq{\Theta}=\oo$.
 
3. $A_1\mlc\ldots\mlc A_n$ is defined by: 
$\Gamma\in \legal{A_1\mlc\ldots\mlc A_n}{}$ iff every move of $\Gamma$ starts with `$i.$' for one of the $i\in\{1,\ldots,n\}$ and,
for each such $i$, $\Gamma^{i.}\in\legal{A_i}{}$; whenever $\Gamma\in\legal{A_1\mlc\ldots\mlc A_n}{}$,  $\win{A_1\mlc\ldots\mlc A_n}{}\seq{\Gamma}=\pp$ iff, for each $i\in\{1,\ldots,n\}$,     
$\win{A_i}{}\seq{\Gamma^{i.}}=\pp$.

4. $A_1\add\ldots\add A_n$ and $A_1\mld\ldots\mld  A_n$
are defined exactly as $A_1\adc\ldots\adc A_n$ and $A_1\mlc\ldots\mlc  A_n$, respectively, only with ``$\pp$" and ``$\oo$" interchanged. 
And $A\mli B$ is defined as $(\gneg A)\mld B$.

5. The infinite $\adc$-conjunction $A_1\adc A_2\adc\ldots$ is defined exactly as $A_1\adc\ldots\adc A_n$, only with ``$i\in\{1,2,\ldots\}$" instead of ``$i\in\{1,\ldots,n\}$". Similarly for the infinite versions of $\add$, $\mlc$, $\mld$. 

6. In addition to the earlier-established meaning, the symbols $\twg$ and $\tlg$ also denote two special --- simplest --- games, defined by $\legal{\twg}{}=\legal{\tlg}{}=\{\emptyrun\}$, $\win{\twg}{}\emptyrun=\pp$ and $\win{\tlg}{}\emptyrun=\oo$. 
\end{definition}

An important operation not explicitly mentioned in Section \ref{ss2} is what is called {\em prefixation}.
This operation takes two arguments: a constant game $A$ and a position $\Phi$ 
 that must 
be a legal position of $A$ (otherwise the operation is undefined), and returns the game $\seq{\Phi}A$.
Intuitively, $\seq{\Phi}A$ is the game playing which means playing $A$ starting (continuing) from position $\Phi$. 
That is, $\seq{\Phi}A$ is the game to which $A$ {\bf evolves} (will be ``{\bf brought down}") after the moves of $\Phi$ have been made. We have already used this intuition when explaining the meaning of choice operations in Section \ref{ss2}: we said that after $\oo$ makes an initial move $i\in\{1,\ldots,n\}$,
 the game 
$A_1\adc\ldots\adc A_n$ continues as $A_i$. What this meant was nothing but that 
$\seq{\oo i}(A_1\adc\ldots\adc A_n)=A_i$.
Similarly, $\seq{\pp i}(A_1\add \ldots\add A_n)=A_i$. Here is the definition of prefixation:

\begin{definition}\label{prfx}
Let $A$ be a constant game and $\Phi$ a legal position of $A$. The game 
$\seq{\Phi}A$ is defined by: $\legal{\seq{\Phi}A}{}=\{\Gamma\ |\ \seq{\Phi,\Gamma}\in\legal{A}{}\}$;
$\win{\seq{\Phi}A}{}\seq{\Gamma}=\win{A}{}\seq{\Phi,\Gamma}$.
\end{definition}
 
The operation $\st$ is somewhat more complex and its definition relies on certain additional conventions. 
We will be referring to (possibly infinite) strings of $0$s and $1$s  as {\bf bit strings}, using  
the letters $w$, $u$ as metavariables for them. The expression $wu$, meaningful only when $w$ is finite, will stand for the concatenation of strings $w$ and $u$. We write $w\preceq u$\label{0preceq} to mean that $w$ is a (not necessarily proper) initial segment of $u$. The letter $\epsilon$ will exclusively stand for the {\bf empty bit string}.

\begin{convention}\label{tree}
By a 
{\bf tree}\label{0tr} we mean a nonempty set $T$ of bit strings,  called {\bf branches}\label{0branch} of the tree, such that, for every 
$w,u$, we have: (a) if $w\in T$ and $u\preceq w$, then  $u\in T$; (b) 
$w0\in T$ iff $w1\in T$; (c) if $w$ is infinite and every finite $u$ with $u\preceq w$ is in $T$, then $w\in T$. Note that $T$ is finite iff every branch of it is finite. 
A {\bf complete branch}\label{0cbranch} of $T$ is a branch $w$ such that no bit string $u$ with $w\preceq u \not= w$ is
in $T$. Finite branches are called {\bf nodes}, and complete finite branches called {\bf leaves}.\end{convention}

\begin{definition}\label{prleg}
We define the notion of a {\bf prelegal position}\label{0prelegal}, together with the function {\em Tree}\label{0treee} that associates a finite tree 
$\tree{}{\Phi}$ with each prelegal position $\Phi$, by the following induction:
\begin{enumerate} 
\item $\emptyrun$ is a prelegal position, and $\tree{}{}=\{\epsilon\}$.
\item $\seq{\Phi,\lambda}$ is a prelegal position  iff $\Phi$ is so and one of the following two conditions is satisfied: 
\begin{enumerate}
\item $\lambda=\oo \col{w}$ for some  leaf $w$ of $\tree{\cal T}{\Phi}$. We call this sort of a labmove $\lambda$ {\bf replicative}.\label{0replm} In this case $\tree{\cal T}{\Phi,\lambda}=\tree{\cal T}{\Phi}\cup\{w0,w1\}$.
\item $\lambda$ is $\oo w.\alpha$ or $\pp w.\alpha$ for some node $w$ of $\tree{\cal T}{\Phi}$ and move $\alpha$. We call this sort of a labmove $\lambda$ {\bf nonreplicative}.\label{0nonrlm} In this case $\tree{\cal T}{\Phi,\lambda}=\tree{\cal T}{\Phi}$.
\end{enumerate}
\end{enumerate}
The terms ``replicative" and ``nonreplicative" also extend from labmoves to moves. 
 When a run $\Gamma$ is infinite, it is considered prelegal iff all of its finite initial segments are so. For such a $\Gamma$, the value of $\tree{}{\Gamma}$ is the smallest tree such that, for every finite initial segment $\Phi$ of $\Gamma$, $\tree{}{\Phi}\subseteq \tree{}{\Gamma}$.

\end{definition}

\begin{convention}\label{succeq}
Let $u$ be a bit string and $\Gamma$ any run. Then
\(\Gamma^{\preceq u}\)\label{susu}
will stand for the result of first removing from $\Gamma$ all labmoves except those that look like
$\xx w.\alpha$ for some  bit string $w$ with $w\preceq u$, and then deleting this sort of prefixes `$w.$' 
in the remaining labmoves, i.e. replacing each remaining labmove $\xx w.\alpha$ (where $w$ is a bit string) by $\xx\alpha$.
Example: If $u=101000\ldots$ and $\Gamma=\seq{\pp \epsilon.\alpha_1,\oo \col{}, \oo 1.\alpha_2, \pp 0.\alpha_3,\oo \col{1},\pp 10.\alpha_4}$, then $\Gamma^{\preceq u}=\seq{\pp\alpha_1, \oo\alpha_2,\pp \alpha_4}$. 
\end{convention}

\begin{definition}\label{legst}
Let $A$ be a constant game. The game \(\st A\) is defined by: 
\begin{enumerate}
\item A position $\Phi$ is in $\legal{\sti A}{}$  iff $\Phi$ is prelegal and, for every leaf $w$ of $\tree{}{\Phi}$, 
 $\Phi^{\preceq w}\in\legal{A}{}$.\vspace{-5pt}
\item As long as  $\Gamma\in\legal{\sti A}{}$, $\win{\sti A}{}\seq{\Gamma}=\pp$ iff $\win{A}{}\seq{\Gamma^{\preceq u}}=\pp$  for every infinite bit string $u$.\footnote{For reasons pointed out on page 39 of \cite{Jap03}, the phrase ``for every infinite bit string $u$" here can be equivalently replaced by ``for every complete branch $u$ of $\tree{}{\Gamma}$". Similarly, in clause 1, ``every leaf $w$ of $\tree{}{\Phi}$" can be replaced by ``every infinite bit string $w$".}  
\end{enumerate}
\noindent Next, we officially reiterate the earlier-given definition of $\intimpl$ by stipulating that $A\intimpl B=_{def}\st A\mli B$.
\end{definition}

\begin{remark}\label{newrem}
Intuitively, a legal run $\Gamma$ of $\st A$ can be thought of as a multiset $Z$ of parallel legal runs of $A$. Specifically,
$Z=\{\Gamma^{\preceq u}\ |$ {\em $u$ is a complete branch of} $\tree{}{\Gamma}\}$, with complete branches of
 $\tree{}{\Gamma}$ thus acting as names for --- or ``representing" --- elements of $Z$. 
In order for $\pp$ to win,  
every run from $Z$ should be a $\pp$-won run of $A$.
The runs from $Z$ typically share some common initial segments and, put together, can be seen as forming a tree of labmoves, with 
$\tree{}{\Gamma}$ --- that we call the {\em underlying tree-structure} of $Z$ --- in a sense presenting the shape of that tree. 
 The meaning of a replicative move $\col{w}$ --- making which is an exclusive privilege of $\oo$ --- is creating in (the evolving) $Z$ two copies of position $\Gamma^{\preceq w}$ out of one. 
 And the meaning of a nonreplicative move $w.\alpha$ is making move $\alpha$ in all positions $\Gamma^{\preceq u}$ of 
(the evolving) $Z$ with $w\preceq u$.
 This is a brutally brief explanation, of course.
The reader may find it very helpful to see Section 13 of \cite{Jap03} 
for detailed explanations and illustrations of the intuitions associated with our $\st$-related formal concepts.\footnote{A few potentially misleading typos have been found in Section 13 (and beyond) of \cite{Jap03}.
The current erratum note is maintained at http://www.csc.villanova.edu/$^\sim$japaridz/CL/erratum.pdf} 
\end{remark}

\section{Not-necessarily-constant games}\label{nncg}

Constant games can be seen as generalized propositions: while propositions in classical logic are just elements 
of $\{\twg,\tlg\}$, constant games are functions from runs to $\{\twg,\tlg\}$.
As we know, however, propositions only offer a very limited expressive power, 
and classical logic needs 
to consider the more general concept of predicates, with propositions being nothing but special --- constant --- cases of predicates. The situation in CL is similar. Our concept of (simply) game generalizes that of constant game in the same sense as the classical concept of predicate generalizes that of proposition.

Let us fix two infinite sets of expressions: the set 
$\{v_1,v_2,\ldots\}$ of {\bf variables} and the set $\{1,2,\ldots\}$ of {\bf constants}. Without loss of generality here we assume that the above collection of constants is exactly the universe of discourse
--- i.e. the set over which the variables range --- in all cases that we consider. By a {\bf valuation} we mean 
a function $e$ that sends each variable $x$ to a constant $e(x)$. In these terms, a classical predicate $P$ can be understood as 
a function that sends each valuation $e$ to a proposition, i.e. constant predicate.   Similarly, what we call a game sends valuations to constant games: 

\begin{definition}\label{ngame}
A {\bf game} is a function $A$ from valuations to constant games. We write $e[A]$ (rather than $A(e)$) to denote the constant game returned by $A$ for valuation $e$. Such a constant game $e[A]$ is said to be an {\bf instance} of $A$. For readability, we often write $\legal{A}{e}$ and $\win{A}{e}$ instead of $\legal{e[A]}{}$ and $\win{e[A]}{}$.
\end{definition}

Just as this is the case with propositions versus predicates, constant games in the sense of Definition \ref{game} will
be thought of as special, constant cases of games in the sense of Definition \ref{ngame}. In particular, each constant game $A'$ is the game $A$ such that, for every valuation $e$,
$e[A]=A'$. From now on we will no longer distinguish between such $A$ and $A'$, so that, if $A$ is a constant game,
it is its own instance, with $A=e[A]$ for every $e$.      
 
We say that a game $A$ {\bf depends on} a variable $x$ iff there are two valuations $e_1,e_2$ that agree on all variables except $x$ such that $e_1[A]\not=e_2[A]$. Constant games thus do not depend on any variables. 
 
Just as the Boolean operations straightforwardly extend from propositions to all predicates, our operations 
$\gneg,\mlc,\mld,\mli,\adc,\add,\st,\intimpl$ extend from constant games to all games. This is done by simply stipulating that $e[\ldots]$ commutes with all of those operations: $\gneg A$ is 
the game such that, for every $e$, $e[\gneg A]=\gneg e[A]$; $A\adc B$ is the game such that,
for every $e$, $e[A\adc B]=e[A]\adc e[B]$; etc. 

To generalize the standard operation of substitution of variables to games, let us agree that by a {\bf term} we mean either 
a variable or a constant; the domain of each valuation $e$ is extended to all terms by stipulating that, for any constant $c$, $e(c)=c$. 

\begin{definition}\label{sov}
Let $A$ be a game, $x_1,\ldots,x_n$ pairwise distinct variables, and $t_1,\ldots,t_n$ any (not necessarily distinct) terms.
The result of {\bf substituting $x_1,\ldots,x_n$ by $t_1,\ldots,t_n$ in $A$}, denoted $A(x_1/t_1,\ldots,x_n/t_n)$, is defined by stipulating that, for every valuation $e$, $e[A(x_1/t_1,\ldots,x_n/t_n)]=e'[A]$, where $e'$ is the valuation for which we have  $e'(x_1)=e(t_1)$, \ldots, $e'(x_n)=e(t_n)$ and, for every variable $y\not\in\{x_1,\ldots,x_n\}$, $e'(y)=e(y)$.
\end{definition}

Intuitively $A(x_1/t_1,\ldots,x_n/t_n)$ is $A$ with $x_1,\ldots,x_n$ remapped to $t_1,\ldots,t_n$, respectively. 
Following the standard readability-improving practice established in the literature for predicates, we will often fix a tuple $(x_1,\ldots,x_n)$ of pairwise distinct variables for a game $A$ and write $A$ as $A(x_1,\ldots,x_n)$. 
It should be noted that when doing so, by no means do we imply that $x_1,\ldots,x_n$ are of all of 
(or only) the variables on which $A$ depends. Representing $A$ in the form $A(x_1,\ldots,x_n)$ sets a context in which we can write $A(t_1,\ldots,t_n)$ to mean the same as the more clumsy expression $A(x_1/t_1,\ldots,x_n/t_n)$. 

In the above terms, we now officially reiterate the earlier-given definitions of the two main quantifier-style operations $\ada$ and $\ade$:  
$\ada xA(x)=_{def}A(1)\adc A(2)\adc A(3)\adc\ldots$ and $\ade xA(x)=_{def}A(1)\add A(2)\add A(3)\add\ldots$.

\section{Computational problems and their algorithmic solvability}\label{icp}

Our games are obviously general enough to model anything that one would call a (two-agent, 
two-outcome) interactive problem.  However, they are a bit too general. There are games where the chances of a player to succeed essentially depend on the relative speed at which its adversary acts. A simple example would be a game where both players have a legal move  in the initial position, and which is won by the player who moves first.  CL
does not want to consider this sort of games meaningful computational problems. Definition 4.2 
of \cite{Jap03} imposes a simple condition on games and calls games satisfying that condition {\bf static}. We are not reproducing that definition here as it is not relevant for our purposes. It should be however mentioned that, according 
to one of the theses on which the philosophy of CL relies, the concept of static games is an adequate formal counterpart of our intuitive concept of ``pure", speed-independent interactive computational problems. All meaningful and reasonable examples of games --- including all elementary games --- are static, and the class of static games is closed under all of the game operations that we have seen (Theorem 14.1 of \cite{Jap03}). 
Let us agree that from now on the term ``{\bf computational problem}", or simply ``{\bf problem}", is a synonym of ``static game".

 Now it is time to explain what {\em computability} of such problems means. The definitions given in this section are semiformal. The omitted technical details are rather standard or irrelevant and can be easily restored by anyone familiar with Turing machines. 
If necessary, the corresponding detailed definitions can be found in Part II of \cite{Jap03}. 
  
\cite{Jap03} defines two models of interactive computation, called the {\em hard-play machine} ({\em HPM}) and the {\em easy-play machine} ({\em EPM}). Both are sorts of Turing machines with the capability of making moves, and have three tapes: the ordinary read/write {\em work tape}, and the read-only {\em valuation} and {\em run} tapes. The 
valuation tape contains a full description of some valuation $e$ (say, by listing the values of $e$ at $v_1,v_2,\ldots$), and its content remains fixed throughout the work of the machine. As for the run tape, it serves as a dynamic input, at any time spelling     
the current position, i.e. the sequence of the (lab)moves made by the two players so far. So, every time one of the players makes a move, that move --- with the corresponding label --- is automatically appended to the content of this tape. 
In the HPM model, there is no restriction on the frequency of environment's moves.
In the EPM model, on the other hand, the machine has full control over the speed of its adversary: the environment can 
(but is not obligated to) make a (one single) move only when the machine explicitly allows it to do so --- the event that we call {\bf granting permission}. The only ``fairness" requirement that such a machine is expected to satisfy is that it should grant permission every once in a while; how long that ``while" lasts, however, is totally up to the machine.  
The HPM and EPM models seem to be two extremes, yet, according to Theorem 17.2 of \cite{Jap03}, they yield the same class of 
winnable static games. The present paper will only deal with the EPM model, so let us take a little closer look at it.

Let $\cal M$ be an EPM. A {\em configuration}\label{y10} of $\cal M$ is defined in the standard way: this is a full description of the (``current") state of the machine, the locations of its three scanning heads
and the contents of its tapes, with the exception that, in order to make finite descriptions of configurations possible, we do not formally include a description of the unchanging 
(and possibly essentially infinite) content of the valuation tape as a part of configuration, but rather account for it in our definition of computation branch as will be seen below. 
The {\em initial configuration} is the configuration where $\cal M$ is in its start state and the work and run tapes are empty. A configuration $C'$ is said to be an {\em $e$-successor} of configuration $C$ in $\cal M$ if, when valuation $e$ is spelled on the valuation tape,  $C'$ can legally follow $C$ in the standard --- standard for multitape Turing machines --- sense, based on the transition function (which we assume to be deterministic) of the machine and accounting for the possibility of nondeterministic updates 
--- depending on what move $\oo$ makes or whether it makes a move at all --- of the content of the run tape when $\cal M$ grants permission. Technically granting permission happens by entering one of the specially designated states called ``permission states".  An {\bf $e$-computation branch}\label{y4} of $\cal M$ is a sequence of configurations of $\cal M$ where the first
configuration is the   initial configuration and every other configuration is an $e$-successor of the previous one.
Thus, the set of all $e$-computation branches captures all possible scenarios (on valuation $e$) corresponding to different behaviors by $\oo$. Such a branch is said to be {\bf fair} iff permission is granted infinitely many times in it.
Each $e$-computation branch $B$ of $\cal M$ incrementally spells --- in the obvious sense --- a run $\Gamma$ on the run tape, which we call the {\bf run spelled by $B$}. Then, for a game $A$ we write ${\cal M}\models_e A$ to mean that, whenever $\Gamma$  is the run spelled by some $e$-computation branch $B$ of $\cal M$ and $\Gamma$ is not $\oo$-illegal, then  branch $B$ is fair and $\win{A}{e}\seq{\Gamma}=\pp$. We write ${\cal M}\models A$ and say that $\cal M$ {\bf computes} ({\bf solves, wins}) $A$ iff ${\cal M}\models_e A$ for every valuation $e$. Finally, we write $\models A$ and say that 
$A$ is {\bf computable} iff there is an EPM $\cal M$ with ${\cal M}\models A$.

\section{The language of $\hint$ and the extended language}\label{s2}

As mentioned, the language of intuitionistic logic can be seen as a fragment of that of CL.
The main building blocks of the language of $\hint$ are infinitely many {\bf problem letters}, or 
{\bf letters} for short, for which we use $P,Q,R,S,\ldots$ as metavariables. 
They are what in classical logic are called `predicate letters', and what CL calls `general letters'.
With each letter is associated a nonnegative integer called its {\em arity}. $\intf$ is one of the letters, with arity $0$.  We refer to it as the {\bf logical letter}, and call all other letters {\bf nonlogical}. The language also contains infinitely many {\em variables} and {\em constants} --- exactly the ones fixed in Section \ref{nncg}. ``{\em Term}" also has the same meaning as before.  An {\bf atom} is $P(t_1,\ldots,t_n)$, where $P$ is an $n$-ary letter and 
the $t_i$ are terms. 
Such an atom is said to be {\em $P$-based}. 
If here each term $t_i$ is a constant, we say that 
$P(t_1,\ldots,t_n)$ is {\bf grounded}. 
A $P$-based atom is $n$-ary, logical, nonlogical etc. iff $P$ is so. When $P$ is $0$-ary, we write $P$ instead of $P()$. {\bf $\hint$-Formulas} are the elements of the smallest class of expressions such that all atoms are $\hint$-formulas and, if $F,F_1,\ldots,F_n$ ($n\geq 2$) are $\hint$-formulas and $x$ is a variable, then 
the following expressions are also $\hint$-formulas: $(F_1)\intimpl (F_2)$, $(F_1)\adc\ldots\adc (F_n)$, $(F_1)\add\ldots\add (F_n)$, $\ada x(F)$, $\ade x(F)$.  Officially there is no negation in the language of $\hint$. Rather, the intuitionistic negation of $F$ is understood as $F\intimpl \intf$. 
In this paper we also employ a more expressive formalism that we call the {\bf extended language}. The latter has the additional connectives $\twg,\tlg,\gneg, \mlc,\mld,\mli,\st$ on top of those of the language of $\hint$, extending the above formation rules by adding the clause that $\twg$, $\tlg$, 
$\neg F$, $(F_1)\mlc\ldots\mlc (F_n)$, $(F_1)\mld\ldots\mld (F_n)$, $(F_1)\mli (F_2)$ and $\st (F)$ are formulas as long as 
$F$, $F_1,\ldots,F_n$ are so. $\twg$ and $\tlg$ count as logical atoms. Henceforth by (simply) ``{\bf formula}", unless otherwise specified, we mean a formula of the extended language. Parentheses will often be omitted in formulas when this causes no ambiguity. With $\ada$ and $\ade$ being {\bf quantifiers}, the definitions of {\bf free} and {\bf bound} occurrences of variables are standard.  

In concordance with a similar notational convention for games on which we agreed in Section \ref{nncg}, 
sometimes a formula $F$ will be represented as $F(x_1,\ldots,x_n)$, where the $x_i$ are pairwise distinct variables. 
When doing so, we do not necessarily mean that each such $x_i$ has a free occurrence in $F$, or that every variable occurring free in $F$ is among $x_1,\ldots,x_n$. In the context set by the above representation, $F(t_1,\ldots,t_n)$ will mean the result of replacing, in $F$, each free occurrence of each $x_i$ ($1\leq i\leq n$) by term $t_i$. In case each $t_i$ is a variable $y_i$, it may be not clear whether $F(x_1,\ldots,x_n)$ or $F(y_1,\ldots,y_n)$ was originally meant to represent $F$ 
in a given context. 
Our disambiguating convention is that the context is set by the expression that was used earlier. That is, when we first mention $F(x_1,\ldots,x_n)$ and only after that 
use the expression $F(y_1,\ldots,y_n)$, the latter should be understood as the result of replacing variables in the former rather than vice versa. 

Let $x$ be a variable, $t$ a term and $F(x)$ a formula. $t$ is said to be {\bf free for $x$ in $F(x)$} iff 
none of the free occurrences of $x$ in $F(x)$ is in the scope of $\ada t$ or $\ade t$.  Of course, when $t$ is a constant, this condition is always satisfied. 
  
An {\bf interpretation} is a function $^*$ that sends each $n$-ary letter $P$ to a static game $^*P = P^*(x_1,\ldots,x_n)$, where the $x_i$ are pairwise distinct variables. This function induces a unique mapping that sends each formula $F$ to a game $F^*$ (in which case we say that $^*$ {\bf interprets} $F$ as $F^*$ and that $F^*$ is the {\bf interpretation of $F$} under $^*$) by stipulating that:\vspace{-6pt} 
\begin{enumerate}
\item Where $P$ is an $n$-ary letter with $^*P=P^*(x_1,\ldots,x_n)$ and $t_1,\ldots,t_n$ are terms, $(P(t_1,\ldots,t_n))^*=P^*(t_1,\ldots,t_n)$.\vspace{-6pt} 
\item $^*$ commutes with all operators: $\twg^*=\twg$, $(F\intimpl G)^*=F^*\intimpl G^*$, $(F_1\mlc\ldots\mlc F_n)^*=F^{*}_{1}\mlc\ldots\mlc F^{*}_{n}$, $(\ada x F)^*=\ada x(F^*)$, etc.\vspace{-6pt} 
\end{enumerate}

When a given formula is represented as $F(x_1,\ldots,x_n)$, we will typically write $F^*(x_1,\ldots,x_n)$ instead of 
$\bigl(F(x_1,\ldots,x_n)\bigr)^*$.

For a formula $F$, we say that an interpretation $^*$ is {\bf admissible for $F$}, or simply {\bf $F$-admissible}, iff the following conditions are satisfied:\vspace{-4pt}

\begin{enumerate}
\item For every $n$-ary letter $P$ occurring in $F$, where $^*P=P^*(x_1,\ldots,x_n)$, the game $P^*(x_1,\ldots,x_n)$ does not depend on any variables that 
are not among $x_1,\ldots,x_n$ but occur (whether free or bound) in $F$.\vspace{-4pt}
\item $\intf^*= B\adc F^{*}_{1}\adc F^{*}_{2}\adc \ldots$, where $B$ is an arbitrary problem and $F_{1},F_{2},\ldots$ is the lexicographic list of all grounded nonlogical atoms of the language.\vspace{-4pt}
\end{enumerate}

Speaking philosophically, an admissible interpretation $^*$ interprets $\intf$ as a ``strongest problem": the interpretation of every grounded atom and hence --- according to Lemma \ref{l6b} --- of every formula is reducible to $\intf^*$, and reducible in a certain uniform, interpretation-independent way. Viewing $\intf^*$ as a resource, it can be seen as a universal resource that allows its owner to solve any problem. Our choice of the dollar notation here is no accident: money is an illustrative  example of an all-powerful resource in the world where everything can be bought. ``Strongest",  ``universal" or ``all-powerful" do not 
necessarily mean ``impossible". So, the intuitionistic negation $F\intimpl\intf$ of $F$ here does not have the traditional ``$F^*$ is absurd" meaning. Rather, it means that $F^*$ (too) is of universal strength. Turing completeness, NP-completeness and similar concepts are good examples of ``being of universal strength".  
$\intf^*$ is what \cite{Jap03} calls a  {\em standard universal problem} of the universe $\seq{F^{*}_{1}, F^{*}_{2},\ldots}$.
Briefly, a {\em universal problem} of a universe (sequence) $\seq{A_1,A_2,\ldots}$ of problems is a problem $U$ such that $\models U\mli A_1\adc A_2\adc\ldots$ and hence $\models U\intimpl A_1\adc A_2\adc\ldots$, intuitively meaning a problem to which each $A_i$ is reducible. For every $B$, the problem $U=B\adc A_1\adc A_2\ldots$ satisfies this condition, and universal problems of this particular form are called {\em standard}. 
Every universal problem $U$ of a given universe can be shown to be equivalent to a 
standard universal problem $U'$ of the same universe, in the sense that $\models U\intimpl U'$ and $\models U'\intimpl U$. And all of the operators of $\hint$ can be shown to preserve equivalence. Hence, restricting universal problems to standard ones does not result in any loss of generality: a universal problem can always be safely assumed to be standard.  See section 23 of \cite{Jap03} for an extended discussion of the philosophy and intuitions associated with universal problems.
Here we only note that interpreting $\intf$ as a universal problem rather than (as one might expect) as 
$\tlg$ yields more generality, for $\tlg$ is nothing but a special, extreme case of a universal problem. {\em Our soundness theorem for $\hint$, of course, continues to hold with $\tlg$ instead of $\intf$}.\vspace{2pt} 

Let $F$ be a formula. We write $\valid F$ and say that $F$ is {\bf valid} iff $\models F^*$ for every $F$-admissible interpretation $^*$. 
For an EPM $\cal E$, we write ${\cal E}\uvalid F$ and say that $\cal E$ is a {\bf uniform solution}
for $F$ iff ${\cal E}\models F^*$ for every $F$-admissible interpretation $^*$. Finally, we write $\uvalid F$ and 
say that $F$ is {\bf uniformly valid} iff there is a uniform solution for $F$. Note that uniform validity automatically implies validity but not vice versa. Yet, these two concepts have been conjectured to be extensionally equivalent
(Conjecture 26.2 of \cite{Jap03}).  

\section{The Gentzen-style axiomatization of $\hint$} \label{shint}
A {\bf sequent} is a pair $\s{G}\Rightarrow K$, where $K$ is an $\hint$-formula and $\s{G}$ is a (possibly empty) finite sequence of $\hint$-formulas. In what follows, $E,F,K$ will be used as metavariables for formulas, and $\s{G},\s{H}$ as metavariables 
for sequences of formulas. We think of sequents as formulas, identifying $\Rightarrow K$ with $K$, 
$F\Rightarrow K$ with $\st F\mli K$ (i.e. $F\intimpl K$), and \(E_1,\ldots,E_n\Rightarrow K\) ($n\geq 2$) with 
\(\st E_1\mlc\ldots\mlc\st E_n\mli K\).\footnote{In order to fully remain within the language of $\hint$, we could understand \(E_1,\ldots,E_n\Rightarrow K\) as $E_1\intimpl(E_2\intimpl\ldots\intimpl(E_n\intimpl K)\ldots)$, which
can be shown to be equivalent to our present understanding. We, however, prefer to read \(E_1,\ldots,E_n\Rightarrow K\) as \(\sti E_1\mlc\ldots\mlc\sti E_n\mli K\) as 
it seems more convenient to work with.} 
This allows us to automatically extend the concepts such as validity, uniform validity, etc.\ from formulas to sequents. A formula $K$ is considered {\bf provable} in $\hint$ iff the sequent $\Rightarrow K$ is so. 

Deductively, logic $\hint$ is given by the following 15 rules. This axiomatization is known (or can be easily shown) to be equivalent to other ``standard" formulations, including Hilbert-style axiomatizations for $\hint$ and/or versions where 
a primitive symbol for negation is present while $\intf$ is absent, or where $\adc$ and $\add$ are strictly binary, or where variables are the only terms of the language.\footnote{That we allow constants is only for technical convenience. This does not really yield a stronger language, as constants behave like free variables and can be thought of as such.} 

Below $\s{G},\s{H}$ are any (possibly empty) sequences of $\hint$-formulas; $n\geq 2$; $1\leq i\leq n$; $x$ is any variable; $E$, $F$, $K$, $F_1$, \ldots, $F_n$, $K_1$, \ldots, $K_n$, $F(x)$, $K(x)$ are any $\hint$-formulas; $y$ is any variable not occurring (whether free or bound) in the conclusion
of the rule; in Left $\ada$ (resp. Right $\ade$), $t$ is any term free for $x$ in $F(x)$ (resp. 
in $K(x)$). 
${\cal P}\concl C$ means ``from premise(s) $\cal P$ conclude $C$". When there are multiple premises in $\cal P$, they are 
separated with a semicolon.\vspace{5pt}

\noindent $\begin{array}{lrcl}
\mbox{\bf Identity}\vspace{6pt} & & \concl & K\Rightarrow K \\ 

\mbox{\bf Domination}\vspace{6pt} & & \concl & \intf\Rightarrow K\\ 

\mbox{\bf Exchange}\vspace{6pt} & \s{G},E,F,\s{H}\Rightarrow K & \concl & 
\s{G},F,E,\s{H}\Rightarrow K\\

\mbox{\bf Weakening}\vspace{6pt} & \s{G}\Rightarrow K & \concl & \s{G},F\Rightarrow K\\

\mbox{\bf Contraction}\vspace{6pt} & \s{G},F,F\Rightarrow K & \concl & 
\s{G},F\Rightarrow K\\

\mbox{\bf Right $\intimpl$}\vspace{6pt} &   \s{G},F\Rightarrow K  & \concl &
\s{G}\Rightarrow F\intimpl K\\

\mbox{\bf Left $\intimpl$}\vspace{6pt} & \s{G},F\Rightarrow K_1; \ \s{H}\Rightarrow K_2   & \concl & 
\s{G},\s{H}, K_2\intimpl F\Rightarrow K_1\\

\mbox{\bf Right $\adc$}\vspace{6pt} & \s{G}\Rightarrow K_1; \ \ldots;\   \s{G}\Rightarrow K_n & \concl & 
\s{G}\Rightarrow K_1\adc\ldots\adc K_n\\

\mbox{\bf Left $\adc$}\vspace{6pt} & \s{G},F_i\Rightarrow K & \concl &
\s{G},F_1\adc\ldots\adc F_n\Rightarrow K\\

\mbox{\bf Right $\add$}\vspace{6pt}   & \s{G}\Rightarrow K_i & \concl & 
\s{G}\Rightarrow K_1\add\ldots\add K_n\\

\mbox{\bf Left $\add$}\vspace{6pt} & \s{G},F_1\Rightarrow K; \ \ldots;  \ \s{G},F_n\Rightarrow K & \concl &
\s{G},F_1\add\ldots\add F_n\Rightarrow K\\

\mbox{\bf Right $\ada$}\vspace{6pt} & \s{G}\Rightarrow K(y) & \concl &
\s{G}\Rightarrow \ada xK(x)\\

\mbox{\bf Left $\ada$}\vspace{6pt} & \s{G},F(t)\Rightarrow K & \concl &
\s{G},\ada xF(x)\Rightarrow K\\

\mbox{\bf Right $\ade$}\vspace{6pt} & \s{G}\Rightarrow K(t) & \concl &
\s{G}\Rightarrow \ade x K(x)\\

\mbox{\bf Left $\ade$}\vspace{6pt} & \s{G},F(y)\Rightarrow K & \concl & 
\s{G},\ade xF(x)\Rightarrow K
\end{array}$

\begin{theorem}\label{main}
{\bf (Soundness:)} If $\hint\vdash S$, then $\valid S$ (any sequent $S$). Furthermore, {\bf (uniform-constructive soundness:)}  there is an effective procedure that takes any $\hint$-proof of any sequent $S$ and constructs a uniform solution for $S$. 
\end{theorem}

\begin{proof} See Section \ref{smain}. \end{proof}

\section{$\clt$-derived validity lemmas}
 
In our proof of Theorem \ref{main} we will need a number of lemmas concerning uniform validity of certain formulas. Some such validity proofs will be given directly in Section \ref{mvl}. But some proofs come ``for free", based on the soundness
 theorem for logic $\clt$ proven in \cite{Japtocl2}. $\clt$ is a propositional logic whose logical atoms are $\twg$ and 
$\tlg$ (but not $\intf$) and whose connectives are $\gneg,\mlc,\mld,\mli,\adc,\add$. It has two sorts of nonlogical atoms,
called {\em elementary} and {\em general}. General atoms are nothing but $0$-ary atoms of our extended language; 
elementary atoms, however, are something not present in the extended language. We refer to the formulas of the language of $\clt$ as {\bf $\clt$-formulas}. 
In this paper, the $\clt$-formulas that do not contain elementary atoms --- including $\twg$ and $\tlg$ that count as such ---  are said to be {\bf general-base}. 
Thus, every general-base formula is a formula of our extended language, and its validity or uniform validity means the same as in Section \ref{s2}.\footnote{These concepts extend to the full language of $\clt$ as well, with interpretations 
required to send elementary atoms to elementary games (i.e. predicates in the classical sense, understood in CL as games that have no nonemty legal runs).} 

Understanding $F\mli G$ as an abbreviation for $\gneg F\mld G$, a {\bf positive occurrence} in a $\clt$-formula is an occurrence that is in the scope of an even number of occurrences of $\gneg$; otherwise the occurrence is {\bf negative}. A {\bf surface occurrence} is an occurrence that is not in the scope of $\adc$ and/or $\add$. The {\bf elementarization} of a $\clt$-formula $F$ is the result of replacing in $F$ every surface occurrence of every subformula of the form $G_1\adc\ldots\adc G_n$ (resp. $G_1\add\ldots\add G_n$) by $\twg$ (resp. $\tlg$), and every positive (resp. negative) surface occurrence of every general atom by $\tlg$ (resp. $\twg$). A $\clt$-formula $F$ is said to be {\bf stable} 
iff its elementarization is a tautology of classical logic. With these conventions, $\clt$ is axiomatized by the following three rules:

\begin{description}
\item[\ \ (a)]  $\vec{H}\concl F$, where $F$ is stable and $\vec{H}$ is the smallest set of formulas such that, 
whenever $F$ has a positive (resp. negative)  surface occurrence of a subformula $G_1\adc\ldots\adc G_n$ (resp. $G_1\add \ldots\add G_n$), for each 
$i\in\{1,\ldots,n\}$, $\vec{H}$ contains the result of replacing that occurrence in $F$ by $G_i$.
\item[\ \ (b)]  $H\concl F$, where $H$ is the result of replacing in $F$ a negative (resp. positive) surface occurrence of a subformula $G_1\adc\ldots\adc G_n$ (resp. $G_1\add\ldots\add G_n$) by $G_i$ for some $i\in\{1,\ldots, n\}$.
\item[\ \ (c)]  $H\concl F$, where $H$ is the result of replacing in $F$ two --- one positive and one negative --- surface occurrences of some general atom by a nonlogical elementary atom that does not occur in $F$.
\end{description}

In this section  $p,q,r,s,t,u,w\ldots$ (possibly with indices) will exlusively stand for nonlogical elementary atoms, and $P,Q,R,S,T,U,W$ (possibly with indices) stand for general atoms. All of these atoms are implicitely assumed to be pairwise distinct in each context.

\begin{convention}\label{con2}
In Section \ref{shint} we started using the notation $\s{G}$ for sequences of formulas. Later we agreed to identify sequences 
of formulas with $\mlc$-conjunctions of those formulas. So, from now on, an underlined expression such as $\s{G}$ will mean 
an arbitrary formula $G_1\mlc \ldots\mlc G_n$ for some $n\geq 0$. Such an expression will always occur in a bigger context such as $\s{G}\mlc F$ or $\s{G}\mli F$; our convention is that, when $n=0$, $\s{G}\mlc F$ and $\s{G}\mli F$ simply mean $F$.

As we agreed that $p,q,\ldots$ stand for elementary atoms and $P,Q,\ldots$ for general atoms, 
$\s{p},\s{q},\ldots$ will correspondingly mean $\mlc$-conjunctions of elementary atoms, and $\s{P},\s{Q},\ldots$ mean 
$\mlc$-conjunctions of general atoms.

We will also be underlining complex expressions such as $F\mli G$, $\ada xF(x)$ or $\st F$. $\s{F\mli G}$ should be understood as $(F_1\mli G_1)\mlc\ldots\mlc (F_n\mli G_n)$, \ $\s{\ada xF(x)}$ as $\ada x F_1(x)\mlc\ldots\mlc\ada xF_n(x)$ (note that only the $F_i$ vary but not $x$), \ $\s{\st F}$ as $\st F_1\mlc\ldots\st F_n$, 
$\st\vspace{1pt}\s{\st F}$ as $\st(\st F_1\mlc\ldots\mlc\st F_n)$, $\s{\st F}\mli \s{F}\mlc G$ as 
$\st F_1\mlc\ldots\st F_n\mli F_1\mlc\ldots\mlc F_n\mlc G$, etc.
\end{convention}
    
The axiomatization of $\clt$ is rather unusual, but it is easy to get a syntactic feel of it once we do a couple of exercises.

\begin{example}\label{exa1}
The following is a $\clt$-proof of $(P\mli Q)\mlc (Q\mli T)\mli (P\mli T)$:

1. $(p\mli q)\mlc (q\mli t)\mli (p\mli t)$ (from $\{\}$ by Rule {\bf (a)}).

2. $(P\mli q)\mlc (q\mli t)\mli (P\mli t)$ (from 1 by Rule {\bf (c)}).

3. $(P\mli Q)\mlc (Q\mli t)\mli (P\mli t)$ (from 2 by Rule {\bf (c)}).

4. $(P\mli Q)\mlc (Q\mli T)\mli (P\mli T)$ (from 3 by Rule {\bf (c)}).
\end{example}

\begin{example}\label{exa2} \ 
Let $n\geq 2$, and let $m$ be the length (number of conjuncts) of both $\s{R}$ and $\s{r}$.  

a) For $i\in\{1,\ldots, n\}$,  the formula of Lemma \ref{l8}(j)
is provable in $\clt$. It follows from $(\s{R}\mli S_i)\mli(\s{R}\mli S_i)$ by Rule {\bf (b)}; the latter follows from 
$(\s{R}\mli s_i)\mli(\s{R}\mli s_i)$ by Rule {\bf (c)}; the latter, in turn, can be derived from 
$(\s{r}\mli s_i)\mli(\s{r}\mli s_i)$ applying Rule {\bf (c)} $m$ times. Finally, $(\s{r}\mli s_i)\mli(\s{r}\mli s_i)$ is 
its own elementarization and is a classical tautology, so it follows from the empty set of premises by Rule {\bf (a)}.

b) The formula of Lemma \ref{l8}(h) is also provable in $\clt$. It is derivable by Rule {\bf (a)} from the set of $n$ premises, easch premise being $(\s{R}\mli S_1)\mlc\ldots\mlc(\s{R}\mli S_n)\mli (\s{R}\mli S_i)$ for some $i\in\{1,\ldots,n\}$. The latter is derivable by Rule {\bf (c)} from  $(\s{R}\mli S_1)\mlc\ldots\mlc (\s{R}\mli s_i)\mlc\ldots\mlc(\s{R}\mli S_n)\mli (\s{R}\mli s_i)$. The latter, in turn, can be derived from 
$(\s{R}\mli S_1)\mlc\ldots\mlc (\s{r}\mli s_i)\mlc\ldots\mlc(\s{R}\mli S_n)\mli (\s{r}\mli s_i)$ applying Rule {\bf (c)} $m$ times. Finally, the latter follows from the empty set of premises by Rule {\bf (a)}. 
\end{example}

Obviously  $\clt$ is decidable. This logic has been proven sound and complete in \cite{Japtocl2}. We only need the soundness part of that theorem restricted to general-base formulas. It sounds as follows:

\begin{lemma}\label{old}
Any general-base $\clt$-provable formula is valid. Furthermore, there is an effective procedure that takes any $\clt$-proof of any such formula $F$ and constructs a uniform solution for $F$. 
\end{lemma}
 
A {\em substitution} is a function $f$ that sends every general atom $P$ of the language of $\clt$ 
to a formula $f(P)$ of the extended language. This mapping extends to all general-base $\clt$-formulas by stipulating that $f$ commutes with all operators: $f(G_1\mli G_2)=f(G_1)\mli f(G_2)$,  
$f(G_1\adc\ldots\adc G_k)=f(G_1)\adc\ldots\adc f(G_k)$, etc.
We say that a formula $G$ is a {\bf substitutional instance} of a general-base $\clt$-formula $F$ iff $G=f(F)$ for some substitution $f$. 
Thus, ``$G$ is a substitutional instance of $F$" means that $G$ has the same form as $F$. 

In the following lemma, we assume $n\geq 2$ (clauses (h),(i),(j)), and $1\leq i\leq n$ (clause (j)). Notice that, for the exception of clause (g), the expressions given below are schemata of formulas rather than formulas, for the lengths of their underlined expressions --- as well as $i$ and $n$ --- may vary. 

\begin{lemma}\label{l8}
All substitutional instances of all formulas given by the following schemata are uniformly valid. Furthermore, there is an effective procedure that takes any particular formula matching a given scheme and constructs 
an EPM that is a uniform solution for all substitutional instances of that formula.\vspace{5pt} 

a) $(\s{R}\mlc P\mlc Q \mlc \s{S} \mli T)\mli(\s{R}\mlc Q\mlc P\mlc \s{S} \mli T)$;

b) $(\s{R} \mli T)\mli(\s{R}\mlc P \mli T)$;

c) $\s{(R\mli S)}\mli (\s{W}\mlc \s{R}\mlc\s{U}\mli \s{W}\mlc\s{S} \mlc\s{U})$;

d) $\bigl(\s{R}\mlc P\mli Q\bigr)\mli \bigl(\s{R}\mli (P\mli Q)\bigr)$;

e) $\bigl(P\mli(Q\mli T)\bigr)\mlc
(\s{R}\mli Q)\mli\bigl(P\mli(\s{R}\mli T)\bigr)$;

f) $\bigl(P\mli(\s{R}\mli Q)\bigr)\mlc
(\s{S}\mlc Q\mli T)\mli
(\s{S}\mlc\s{R}\mlc P\mli T)$;

g) $(P\mli Q)\mlc (Q\mli T)\mli (P\mli T)$;

h) $(\s{R}\mli S_1)\mlc\ldots\mlc(\s{R}\mli S_n)\mli (\s{R}\mli S_1\adc\ldots\adc S_n)$;

i) $(\s{R}\mlc S_1\mli T)\mlc\ldots\mlc(\s{R}\mlc S_n\mli T)\mli \bigl((\s{R}\mlc (S_1\add\ldots\add S_n)\mli T\bigr)$;

j) $(\s{R}\mli S_i)\mli(\s{R}\mli S_1\add\ldots\add S_n)$.
\end{lemma}

\begin{proof} 
In order to prove this lemma, it would be sufficient to show that all formulas given by the above schemata are provable in $\clt$. Indeed, if we succeed in doing so, then an effective procedure whose existence is claimed in the lemma could be designed to work as follows: First, the procedure finds a $\clt$-proof of a given formula $F$. Then, based on that proof and using the procedure whose existence is stated in Lemma \ref{old}, it finds a uniform solution $\cal E$ for that formula. It is not hard to see that the same $\cal E$ will automatically be a uniform solution for every substitutional instance of $F$ as well. 

The $\clt$-provability of the formulas given by clauses (g), (h) and (j) has been verified in Examples \ref{exa1} and \ref{exa2}. A similar verification for the other clauses is left as an easy syntactic exercise for the reader. 
\end{proof}

\section{Closure lemmas}

In this section we let $n$ range over natural numbers (including $0$), $x$ over any variables, $F,E,G$ (possibly with subscripts) over any formulas, and $\cal E$, $\cal C$, $\cal D$ (possibly with subscripts) over any EPMs. Unless otherwise specified, these 
metavariables are assumed to be universally quantified in a context. The expression \{{\em EPMs}\} stands for the set of all EPMs.

\begin{lemma}\label{l10}
If $\uvalid F$, then $\uvalid \st F$.
Furthermore, there is an effective function $h: \ \{\mbox{EPMs}\}\longrightarrow\{\mbox{EPMs}\}$ 
such that, for any ${\cal E}$ and $F$, 
if ${\cal E}\uvalid F$, then $h({\cal E})\uvalid \st F$. 
\end{lemma}

\begin{proof} According to Proposition 21.2 of \cite{Jap03}, there is an effective function $h: \ \{\mbox{EPMs}\}\longrightarrow\{\mbox{EPMs}\}$  such that, for any EPM 
$\cal E$ and any static game $A$, if ${\cal E}\models A$, then $h({\cal E})\models \st A$. We now claim that 
if ${\cal E}\uvalid F$, then $h({\cal E})\uvalid \st F$. Indeed, assume ${\cal E}\uvalid F$. Consider any $\st F$-admissible interpretation $^*$. Of course, the same interpretation is also $F$-admissible. Hence, ${\cal E}\uvalid F$ implies 
${\cal E}\models F^*$. But then, by the known behavior of $h$, we have $h({\cal E})\models \st F^*$. Since $^*$ was arbitrary, we conclude that $h({\cal E})\uvalid \st F$.
\end{proof}

\begin{lemma}\label{l10a}
If $\uvalid F$, then $\uvalid \ada x F$.
Furthermore, there is an effective function $h: \ \{\mbox{EPMs}\}\longrightarrow\{\mbox{EPMs}\}$ 
such that, for any ${\cal E}$, $x$ and $F$, 
if ${\cal E}\uvalid F$, then $h({\cal E})\uvalid \ada x F$. 
\end{lemma}

\begin{proof} Similar to the previous lemma, only based on Proposition 21.1 of \cite{Jap03} instead of 21.2.
\end{proof}

\begin{lemma}\label{l1a} 
If $\uvalid F$ and $\uvalid F\mli E$, then $\uvalid E$. 
Furthermore, there is an effective function $h: \ \{\mbox{EPMs}\}\times\{\mbox{EPMs}\}\longrightarrow\{\mbox{EPMs}\}$ 
such that, for any ${\cal E}$, $\cal C$, $F$ and $E$, 
if ${\cal E}\uvalid F$  and 
${\cal C}\uvalid F\mli E$, then $h({\cal E},{\cal C})\uvalid E$. 
\end{lemma}

\begin{proof} According to Proposition 21.3 of \cite{Jap03}, there is an effective function $g$ that takes any HPM $\cal H$
and  EPM 
${\cal E}$ and returns an EPM ${\cal C}$ such that, for any static games $A$,$B$ and any valuation $e$, if ${\cal H}\models_e A$ and ${\cal E}\models_e A\mli B$, then ${\cal C}\models_e B$. Theorem 17.2 of \cite{Jap03}, which establishes equivalence between EPMs and HPMs in a constructive sense, allows us to assume that $\cal H$ is an EPM rather than HPM here. More precisely, we can 
routinely convert $g$ into an (again effective) function $h: \ \{\mbox{EPMs}\}\times\{\mbox{EPMs}\}\longrightarrow\{\mbox{EPMs}\}$ such that,
for any static games $A$,$B$, valuation $e$ and EPMs ${\cal E}$ and ${\cal C}$, 
\begin{equation}\label{oct9}\mbox{\em if ${\cal E}\models_e A$ and ${\cal C}\models_e A\mli B$, then $h({\cal E},{\cal C})\models_e B$.}
\end{equation}
We claim that the above function $h$ behaves as our lemma states. 
Assume ${\cal E}\uvalid F$  and 
${\cal C}\uvalid F\mli E$, and consider an arbitrary valuation $e$ and an arbitrary $E$-admissible interpretation $^*$.
Our goal is to show that $h({\cal E},{\cal C})\models_e E^*$, which obviously means the same as 
\begin{equation}\label{oct9a}
h({\cal E},{\cal C})\models_e e[E^*].
\end{equation}
We define the new interpretation $^\dagger$ by stipulating that, for every $n$-ary letter $P$ with $P^*= P^*(x_1,\ldots,x_n)$,  $P^\dagger$ is the game $P^\dagger (x_1,\ldots,x_n)$ such that, for any tuple $c_1,\ldots,c_n$ of constants, $P^\dagger(c_1,\ldots,c_n)=e[P^*(c_1,\ldots,c_n)]$. Unlike $P^*(x_1,\ldots,x_n)$ that may depend on some ``hidden" variables (those that are not among $x_1,\ldots,x_n$),  obviously 
$P^\dagger (x_1,\ldots,x_n)$ does not depend on any variables other that $x_1,\ldots,x_n$. This makes $^\dagger$ admissible
for any formula, including $F$ and $F\mli E$. Then our assumptions ${\cal E}\uvalid F$  and 
${\cal C}\uvalid F\mli E$ imply  ${\cal E}\models_e F^\dagger$  and 
${\cal C}\models_e F^\dagger\mli E^\dagger$. Consequently, by  
(\ref{oct9}), $h({\cal E},{\cal C})\models_e E^\dagger$,
i.e. $h({\cal E},{\cal C})\models_e e[E^\dagger]$. Now, with some thought, we can see that $e[E^\dagger]=e[E^*]$. Hence (\ref{oct9a}) is true. 
\end{proof}

\begin{lemma}\label{l1} {\bf (Modus ponens)}
If $\uvalid F_1$, \ldots, $\uvalid F_n$ and $\uvalid F_1\mlc\ldots\mlc F_n\mli E$, then $\uvalid E$.
Furthermore, 
there is an effective function 
$h: \ \{\mbox{EPMs}\}^{n+1}\longrightarrow\{\mbox{EPMs}\}$ 
such that, for any EPMs ${\cal E}_1$, \ldots, ${\cal E}_n$, $\cal C$ and any formulas $F_1,\ldots, F_n,E$, 
if ${\cal E}_1\uvalid F_1$, \ldots, ${\cal E}_n\uvalid F_n$  and 
${\cal C}\uvalid F_1\mlc\ldots\mlc F_n\mli E$, then $h({\cal E}_1,\ldots,{\cal E}_n, {\cal C})\uvalid E$. Such a function, in turn, can be effectively constructed for each particular $n$. 
\end{lemma}

\begin{proof} In case $n=0$, $h$ is simply the identity function $h({\cal C})={\cal C}$. In case $n=1$, $h$ is the function whose existence is stated in Lemma \ref{l1a}. Below we will construct $h$ for  
case $n=2$. It should be clear how to generalize that construction to 
any greater $n$. 

Assume ${\cal E}_1\uvalid F_1$, ${\cal E}_2\uvalid F_2$  and 
${\cal C}\uvalid F_1\mlc F_2\mli E$. 
By Lemma \ref{l8}(d), $(F_1\mlc F_2\mli E)\mli (F_1\mli(F_2\mli E))$ has a uniform solution. 
Lemma \ref{l1a} allows us to combine that solution with ${\cal C}$ and find a uniform solution ${\cal D}_1$ for $F_1\mli(F_2\mli E)$. Now applying 
Lemma \ref{l1a} to ${\cal E}_1$ and ${\cal D}_1$, we find a uniform solution ${\cal D}_2$ for $F_2\mli E$. Finally,
 applying the same lemma to ${\cal E}_2$ and ${\cal D}_2$, we find a 
uniform solution $\cal D$ for $E$. Note that $\cal D$ does not depend on $F_1,F_2,E$, and that we constructed $\cal D$ in an effective way from the arbitrary ${\cal E}_1$, ${\cal E}_2$ and ${\cal C}$. Formalizing this construction yields 
function $h$ whose existence is claimed by the lemma.
\end{proof}

\begin{lemma}\label{l1c} {\bf (Transitivity)} 
If $\uvalid F\mli E$ and $\uvalid E\mli G$, then $\uvalid F\mli G$. 
Furthermore, there is an effective function $h: \ \{\mbox{EPMs}\}\times\{\mbox{EPMs}\}\longrightarrow\{\mbox{EPMs}\}$ 
such that, for any ${\cal E}_1$, ${\cal E}_2$, $F$, $E$ and $G$, 
if ${\cal E}_1\uvalid F\mli E$  and 
${\cal E}_2\uvalid E\mli G$, then $h({\cal E}_1,{\cal E}_2)\uvalid F\mli G$. 
\end{lemma}

\begin{proof} Assume ${\cal E}_1\uvalid F\mli E$ and ${\cal E}_2\uvalid E\mli G$. By Lemma \ref{l8}(g), we also have 
${\cal C}\uvalid (F\mli E)\mlc (E\mli G)\mli (F\mli G)$ for some (fixed) ${\cal C}$.  Lemma \ref{l1} allows us to combine the three uniform solutions and construct a uniform solution $\cal D$ for $F\mli G$. 
\end{proof}

\section{More validity lemmas}\label{mvl}

As pointed out in  Remark 16.4 of \cite{Jap03}, when trying to show that a given EPM wins a given game, it is always safe to assume that the runs that the machine generates are never $\oo$-illegal, i.e. that the environment never makes an illegal move, for if it does, the machine automatically wins. This assumption, that we call the {\bf clean environment assumption}, will always be implicitly present in our winnability proofs.

We will often employ a uniform solution 
for $P\mli P$ called the {\bf copy-cat strategy} (${\cal CCS}$). This strategy, that we already saw in Section \ref{ss2}, consists in mimicking, in the antecedent, the moves made by the 
environment in the consequent, and vice versa. More formally, the algorithm that ${\cal CCS}$ follows is an infinite loop, on every iteration of which ${\cal CCS}$ keeps granting permission until the environment 
makes a move $1.\alpha$ (resp. $2.\alpha$), to which 
the machine responds by the move $2.\alpha$ (resp. $1.\alpha$). As shown in the proof of Proposition 22.1 of \cite{Jap03},
this strategy guarantees success in every game of the form $A\mld \gneg A$ and hence 
$A\mli A$. An important detail is that $\cal CCS$ never looks at the past history of the game, 
i.e. the movement of its scanning head on the run tape is exclusively left-to-right. This guarantees that, even if the original game was something else and it only evolved to $A\mli A$ later as a result of making a series of moves, switching to 
the $\cal CCS$ after the game has been brought down to $A\mli A$ guarantees success no matter what happened in the past. 
    
Thgroughout this section, $F$, $G$, $E$, $K$ (possibly with indices and attached tuples of variables) range over formulas, $x$ and $y$ over variables, $t$ over terms, $n$ over nonnegative integers, $w$ over bit strings, and $\alpha,\gamma$ over moves. These (meta)variables are assumed to be universally quantified in a context unless otherwise specified. In accordance with our earlier convention, $\epsilon$ means the empty string, so that, say, 
`$1.\epsilon.\alpha$' is the same as `$1..\alpha$'.

\begin{lemma}\label{l6a}
$\uvalid \st F\mli F$. Furthermore, there is an EPM $\cal E$ such that, for any $F$, \ ${\cal E}\uvalid \st F\mli F$.
\end{lemma}

\begin{proof} The idea of a uniform solution $\cal E$ for $\st F\mli F$ is very simple: just act as $\cal CCS$, never making any replicative moves in the antecedent and pretending that the latter is $F$ rather than (the stronger) $\st F$.
The following formal description of the interactive algorithm that $\cal E$ follows is virtually the same as that of $\cal CCS$, with the only difference that the move prefix `$1.$' is replaced by `$1.\epsilon.$' everywhere.\vspace{5pt}

{\bf Procedure} LOOP: Keep granting permission until the environment makes a move $1.\epsilon.\alpha$ or $2.\alpha$; 
in the former case make the move $2.\alpha$, and in the latter case make the move $1.\epsilon.\alpha$; then  repeat LOOP.\vspace{5pt} 

Fix an arbitrary valuation $e$, interpretation $^*$ and $e$-computation branch $B$ of $\cal E$. Let $\Theta$ be the run spelled by $B$. From the description of LOOP it is clear that permission will be granted  infinitely many times in $B$, so this branch is fair. Hence, in order to show that $\cal E$ wins the game, it would suffice to show 
that $\win{\sti F^*\mli F^*}{e}\seq{\Theta}=\pp$. 

Let $\Theta_i$ denote the initial segment of $\Theta$ consisting of the (lab)moves made by the players by the beginning of the $i$th iteration of LOOP in $B$ (if such an iteration exists). By induction on $i$, based on the clean environment 
assumption and applying a routine analysis of the the behavior of LOOP and the definitions of the relevant game operations,
one can easily find that

\[\begin{array}{l}\label{oct3}
\mbox{a) } \Theta_i\in\legal{\sti F^*\mli F^*}{e};\\
\mbox{b) } \Theta_{i}^{1.\epsilon.}=\rneg\Theta_{i}^{2.}\ ;\\
\mbox{c) } \mbox{\em All of the moves in $\Theta_{i}^{1.}$ have the prefix `$\epsilon.$'.} 
\end{array}\] 

If LOOP is iterated infinitely many times, then the above obviously extends from $\Theta_i$ to $\Theta$, because 
every initial segment of $\Theta$ is an initial segment of some $\Theta_i$. And if LOOP is iterated only a finite number $m$ of times, then $\Theta=\Theta_m$. This is so because the environment cannot  make a move $1.\epsilon.\alpha$ or $2.\alpha$ during the $m$th iteration (otherwise there would be a next iteration), and any other move would violate the clean environment assumption; and, as long as the environment does not move during a given iteration, neither does the machine.
Thus, no matter whether LOOP is iterated a finite or infinite number of times, we have:

\begin{equation}\label{oct3a}
\begin{array}{l}
\mbox{a) } \Theta\in\legal{\sti F^*\mli F^*}{e};\\
\mbox{b) } \Theta^{1.\epsilon.}=\rneg\Theta^{2.}\ ;\\
\mbox{c) } \mbox{\em All of the moves in $\Theta^{1.}$ have the prefix `$\epsilon.$'.} 
\end{array} 
\end{equation} 

Since $\Theta\in\legal{\sti F^*\mli F^*}{e}$, in order to show that 
$\win{\sti F^*\mli F^*}{e}\seq{\Theta}=\pp$, by the definition of $\mli$, it would suffice to show that either 
$\win{F^*}{e}\seq{\Theta^{2.}}=\pp$ or $\win{\gneg \sti F^*}{e}\seq{\Theta^{1.}}=\pp$.
 So, assume $\win{F^*}{e}\seq{\Theta^{2.}}\not=\pp$, i.e. $\win{F^*}{e}\seq{\Theta^{2.}}=\oo$, i.e. 
$\win{\gneg F^*}{e}\seq{\rneg\Theta^{2.}}=\pp$.
 Then, by clause (b) of (\ref{oct3a}),
$\win{\gneg F^*}{e}\seq{\Theta^{1.\epsilon.}}=\pp$, i.e. 
$\win{F^*}{e}\seq{\rneg\Theta^{1.\epsilon.}}=\oo$, i.e.
$\win{F^*}{e}\seq{(\rneg\Theta^{1.})^{\epsilon.}}=\oo$. 
Pick any infinite bit string $w$. In view of clause (c) of (\ref{oct3a}), 
we obviously have 
$(\rneg\Theta^{1.})^{\epsilon.}=(\rneg\Theta^{1.})^{\preceq w}$. Hence
$\win{F^*}{e}\seq{(\rneg\Theta^{1.})^{\preceq w}}=\oo$. But this, by the definition of $\st$,
implies    $\win{\sti F^*}{e}\seq{\rneg\Theta^{1.}}=\oo$. The latter, in turn, can be rewritten as the desired 
$\win{\gneg \sti F^*}{e}\seq{\Theta^{1.}}=\pp$.

Thus, we have shown that $\cal E$ wins $\st F^*\mli F^*$. Since $^*$ was arbitrary and the work of $\cal E$ did 
not depend on it, we conclude that ${\cal E}\uvalid \st F\mli F$.
\end{proof}

In the subsequent constructions found in this section, $^*$ will always mean an arbitrary but fixed interpretation admissible for the formula whose uniform validity we are trying to prove. Next, $e$ will always mean 
an arbitrary but fixed valuation --- specifically, the valuation spelled on the valuation tape of the machine under question. For readability, we will usually omit the $e$ parameter when it is irrelevant. 
Also,
having already seen one example, in the remaining uniform validity proofs we will typically limit ourselves to just constructing interactive algorithms, 
leaving the (usually routine) verification of  their correctness to the reader. An exception will be the proof of Lemma 
\ref{l5} where, due to the special complexity of the case, correctness verification will be done even more rigorously than 
we did this in the proof of Lemma \ref{l6a}.

\begin{lemma}\label{l4}
$\uvalid \st(F\mli G)\mli(\st F\mli \st G)$. Moreover, there is an EPM $\cal E$ such that, for every $F$ and $G$, \ ${\cal E}\uvalid \st(F\mli G)\mli(\st F\mli \st G)$.
\end{lemma}

\begin{proof} 
A relaxed description of a uniform solution $\cal E$ 
for $\st(F\mli G)\mli(\st F\mli \st G)$
is as follows.
In $\st(F^*\mli G^*)$ and $\st F^*$ the machine is making exactly the same replicative moves (moves of the form $\col{w}$) as the environment is making in $\st G^*$. This ensures that the tree-structures of the three $\st$-components of the game are identical, and now all the machine needs for a success is to win the game 
$(F^*\mli G^*)\mli(F^*\mli G^*)$ within each branch of those trees. This can be easily achieved by applying copy-cat methods to the two occurrences of $F$ and the two occurrences of $G$. 

In precise terms, the strategy that $\cal E$ follows is described by the following interactive algorithm.\vspace{5pt}

{\bf Procedure} LOOP: Keep granting permission until the adversary makes a move $\gamma$. Then:  

If $\gamma=2.2.\col{w}$, then make the moves $1.\col{w}$ and $2.1.\col{w}$, and repeat LOOP;

If $\gamma=2.2.w.\alpha$ (resp. $\gamma=1.w.2.\alpha$), then  make the move $1.w.2.\alpha$ (resp. $2.2.w.\alpha$), and repeat LOOP; 

If $\gamma=2.1.w.\alpha$ (resp. $\gamma=1.w.1.\alpha$), then make the move $1.w.1.\alpha$ (resp. $2.1.w.\alpha$), and repeat LOOP.\vspace{-4pt}
\end{proof}

\begin{lemma}\label{l4a} 
$\uvalid \st F_1\mlc\ldots\mlc \st F_n\mli \st (F_1\mlc\ldots\mlc F_n)$. Furthermore, there is an effective procedure that takes any particular value of $n$ and constructs an EPM $\cal E$ such that, for any $F_1,\ldots,F_n$, \ ${\cal E}\uvalid \st F_1\mlc\ldots\mlc \st F_n\mli \st (F_1\mlc\ldots\mlc F_n)$.

\end{lemma}

\begin{proof} The idea of a uniform solution here is rather similar to the one in the proof of Lemma \ref{l4}. Here is the algorithm:\vspace{5pt}

{\bf Procedure} LOOP: Keep granting permission until the adversary makes a move $\gamma$. Then: 

If $\gamma=2.\col{w}$, then make the $n$ moves $1.1.\col{w},\ldots,1.n.\col{w}$, and repeat LOOP;

If $\gamma=2.w.i.\alpha$ (resp. $\gamma=1.i.w.\alpha$) where $1\leq i\leq n$, then make the move $1.i.w.\alpha$ 
(resp. $2.w.i.\alpha$), and repeat LOOP.\vspace{-4pt}
\end{proof}

\begin{lemma}\label{l6c}
$\uvalid \st F\mli\st F\mlc\st F$. Furthermore, there is an EPM $\cal E$ such that, for any $F$, \ ${\cal E}\uvalid 
\st F\mli\st F\mlc\st F$.

\end{lemma}

\begin{proof} The idea of a winning strategy here is to first replicate the antecedent turning it into something ``essentially the same"\footnote{Using the notation $\circ$ introduced in Section 13 of \cite{Jap03}, in precise terms this ``something" is 
$\sti(F^*\circ F^*)$.} as $\st F^*\mlc \st F^*$, and then switch to a strategy that is ``essentially the same as" the ordinary copy-cat strategy. 
Precisely, here is how $\cal E$ works: it makes the move $1.\col{\epsilon}$ (replicating the antecedent), after which it follows the following algorithm:\vspace{7pt}

{\bf Procedure} LOOP: Keep granting permission until the adversary makes a move $\gamma$. Then:

If $\gamma=1.0\alpha$ (resp. $\gamma=2.1.\alpha$), then make the move $2.1.\alpha$ (resp. $1.0\alpha$), and repeat LOOP;

If $\gamma=1.1\alpha$ (resp. $\gamma=2.2.\alpha$), then make the move $2.2.\alpha$ (resp. $1.1\alpha$), and repeat LOOP;

If $\gamma=1.\epsilon.\alpha$, then make the  moves $2.1.\epsilon.\alpha$ and $2.2.\epsilon.\alpha$, and repeat LOOP.
\end{proof}

Remember from Section \ref{nncg} that, when $t$ is a constant, $e(t)=t$. 

\begin{lemma}\label{l6b}
For any $\hint$-formula $K$, $\uvalid \st\intf\mli K$.
Furthermore, there is an effective procedure that takes any $\hint$-formula $K$ and constructs a uniform solution for 
$\st\intf \mli K$.
\end{lemma}

\begin{proof} We construct an EPM $\cal E$ and verify that it is a uniform solution for $\st\intf\mli K$; both the construction and the verification will be done by induction on the complexity of $K$. 
The goal in each case is to show that $\cal E$ generates a $\pp$-won run of  
$e[(\st\intf\mli K)^*]=\st e[\intf^*]\mli e[K^*]$ which, according to our convention, 
we may write with ``$e$" omitted.  

{\em Case 1:} $K=\intf$. This case is taken care of by 
Lemma \ref{l6a}. 

{\em Case 2:} $K$ is a $k$-ary nonlogical atom $P(t_1,\ldots,t_k)$. Let $c_1,\ldots,c_k$ be the constants 
$e(t_1),\ldots,e(t_k)$, respectively. Evidently in this case $e[K^*]=e[P^*(c_1,\ldots,c_k)]$, so, the game for which 
$\cal E$ needs to generate a winning run is 
$\st \intf^*\mli P^*(c_1,\ldots,c_k)$.  Assume $(P(c_1,\ldots,c_k))^*$ is conjunct \#$m$ of 
(the infinite $\adc$-conjunction) $\intf^*$.  We define $\cal E$ to be the 
EPM that acts as follows. At the beginning, if necessary (i.e. unless all $t_i$ are constants), it reads the valuation tape to find $c_1,\ldots,c_k$. Then, using this information,  it finds the above number $m$ and makes the move 
`$1.\epsilon.m$', which can be seen\footnote{See Proposition 13.8 of \cite{Jap03}.} to bring the game down to
 $\st P^*(c_1,\ldots,c_k)\mli P^*(c_1,\ldots,c_k)$. After this move, $\cal E$ switches to the strategy whose existence is stated in  Lemma \ref{l6a}. 

{\em Case 3:} $K=\st E\mli F$.  By Lemma \ref{l8}(b), there is a uniform solution for 
 $(\st\intf\mli F)\mli(\st\intf\mlc\st E\mli F)$. Lemmas \ref{l8}(d)  and \ref{l1c} allow us to convert the latter into 
a uniform solution $\cal D$ for $(\st\intf\mli F)\mli \bigl(\st\intf\mli(\st E\mli F)\bigr)$. By the induction hypothesis, there is also a uniform solution ${\cal C}$ for $\st\intf\mli F$. Applying Lemma \ref{l1} to $\cal C$ and $\cal D$,
we find a uniform solution $\cal E$ for $  \st\intf\mli(\st E\mli F)$.

{\em Case 4:} $K=E_1\add\ldots\add E_n$. By the induction hypothesis, we know how to construct an EPM ${\cal C}_1$ with ${\cal C}_1\uvalid
\st\intf\mli E_1$. Now we define $\cal E$ to be the EPM that first makes the move $2.1$, and then
plays the rest of the game as ${\cal C}_1$ would play. $\cal E$ can be seen to be successful because its initial move $2.1$ brings 
$(\st \intf\mli K)^*$ down to  $(\st\intf\mli E_{1})^{*}$.

{\em Case 5:} $K=E_1\adc\ldots\adc E_n$. By the induction hypothesis, for each $i$ with $1\leq i\leq n$ we have an EPM ${\cal C}_i$ with ${\cal C}_i\uvalid
\st\intf\mli E_i$. We define $\cal E$ to be the EPM that acts as follows. At the beginning, $\cal E$ keeps granting permission until the adversary makes a move. The clean environment assumption guarantees that this move should be $2.i$ for some $1\leq i\leq n$. It brings 
$(\st\intf\mli E_1\adc\ldots\adc E_n)^*$ down to $(\st\intf\mli E_i)^*$. If and after such a move $2.i$ is made, 
$\cal E$ continues the rest of the play as ${\cal C}_i$. 

{\em Case 6:} $K=\ade xE(x)$. By the induction hypothesis, there is an EPM ${\cal C}_1$ with ${\cal C}_1\uvalid
\st\intf\mli E(1)$. Now we define $\cal E$ to be the EPM that first makes the move $2.1$, and then
plays the rest of the game as ${\cal C}_1$.  $\cal E$ can be seen to be successful because its initial move $2.1$ 
brings  
$\bigl(\st \intf\mli \ade xE(x)\bigr)^*$ down to  $\bigl(\st\intf\mli E(1)\bigr)^*$.

{\em Case 7:} $K=\ada xE(x)$. By the induction hypothesis, for each constant $c$ there is (and can be effectively found) an EPM ${\cal C}_c$ with ${\cal C}_c\uvalid
\st\intf\mli E(c)$. Now we define $\cal E$ to be the EPM that acts as follows. At the beginning, $\cal E$ keeps granting permission until the adversary makes a move. By the clean environment assumption, such a move 
should be $2.c$ for some constant $c$.  This move can be seen to  bring 
$\bigl((\st\intf\mli \ada xE(x))\bigr)^*$ down to $\bigl(\st\intf\mli E(c)\bigr)^*$.  
If and after the above move $2.c$ is made, $\cal E$ plays the rest of the game as ${\cal C}_c$.\vspace{-4pt}
\end{proof}

\begin{lemma}\label{l11} 

Assume $n\geq 2$, $1\leq i\leq n$, and $t$ is a term free for $x$ in $G(x)$.  Then the following uniform validities hold. Furthermore, in each case there is an effective procedure that takes any particular values of $n$, $i$, $t$  
and constructs an EPM which is a uniform solution for the corresponding formula no matter what the values of $F_1,\ldots,F_n$ and/or $G(x)$ (as long as $t$ is free for $x$ in $G(x)$) are.\vspace{5pt}

a) $\uvalid \st(F_1\adc\ldots\adc F_n)\mli \st F_i$; 

b) $\uvalid \st\ada xG(x) \mli \st G(t)$; 

c) $\uvalid  \st(F_1\add\ldots\add F_n)\mli \st F_1\add\ldots\add\st F_n$; 

d) $\uvalid \st\ade xG(x) \mli \ade x\st G(x)$. 

\end{lemma}

\begin{proof} Below come winning strategies for each case.

{\em Clause (a):} Make the move `$1.\epsilon.i$'. This brings the game down to $\st F_{i}^{*}\mli \st F_{i}^{*}$. Then switch to 
$\cal CCS$.

{\em Clause (b):} Let $c=e(t)$. Read $c$ from the valuation tape if necessary, i.e., if $t$ is a variable (otherwise, simply $c=t$). Then make the move `$1.\epsilon.c$'. This brings the game down to $\st G^*(c)\mli \st G^*(c)$.
Now, switch to $\cal CCS$.

{\em Clause (c):} Keep granting permission until the adversary makes a move `$1.\epsilon.j$' ($1\leq j\leq n$), 
bringing the game down to 
$\st F_{j}^{*}\mli \st F_{1}^{*}\add\ldots\add\st F_{n}^{*}$. If and after such a move is made (and if not, a win is automatically guaranteed), make the move `$2.j$', which brings the game down to $\st F_{j}^{*}\mli \st F_{j}^{*}$.
Finally, switch to $\cal CCS$.

{\em Clause (d):} Keep granting permission until the adversary makes the move `$1.\epsilon.c$' for some constant $c$. This brings the game down to 
$\st G^*(c)\mli \ade x\st G^*(x)$.  Now make the move `$2.c$', which brings the game down to $\st G^*(c)\mli \st G^*(c)$.
Finally, switch to $\cal CCS$.\vspace{-4pt}
\end{proof}

\begin{lemma}\label{oct5a}
$\uvalid \ada x\bigl(F(x)\mli G(x)\bigr)\mli \bigl(\ada xF(x)\mli\ada xG(x)\bigr)$. Furthermore, there is an EPM $\cal E$ such that, 
for any  $F(x)$ and $G(x)$, \  
${\cal E}\uvalid \ada x\bigl(F(x)\mli G(x)\bigr)\mli \bigl(\ada xF(x)\mli\ada xG(x)\bigr)$.
\end{lemma}

\begin{proof} 
Strategy: Wait till the environment makes the move `$2.2.c$' for some constant $c$. This brings the 
$\ada xG^*(x)$ component down to $G^*(c)$ and hence the entire game to $\ada x\bigl(F^{*}(x)\mli G^*(x)\bigr)\mli \bigl(\ada xF^{*}(x)\mli G^*(c)\bigr)$.
Then make the same move $c$ in the antecedent and in $\ada xF^{*}(x)$, i.e. make the moves 
`$1.c$' and `$2.1.c$'. The game will be brought down to $\bigl(F^{*}(c)\mli G^*(c)\bigr)\mli \bigl(F^{*}(c)\mli G^*(c)\bigr)$. Finally, switch to $\cal CCS$.\vspace{-4pt} 
\end{proof}

\begin{lemma}\label{oct5d}
$\uvalid \ada x\bigl(F_1(x)\mlc\ldots\mlc F_n(x)\mlc E(x)\mli G(x)\bigr)\mli \bigl(\ada xF_1(x)\mlc\ldots\mlc\ada xF_n(x)\mlc\ade xE(x)\mli\ade xG(x)\bigr)$. Furthermore, there is an effective procedure that takes any particular value of $n$ and constructs an EPM $\cal E$ such that, 
for any $F_1(x)$, \ldots, $F_n(x)$, $E(x)$, $G(x)$, \ 
${\cal E}\uvalid \ada x\bigl(F_1(x)\mlc\ldots\mlc F_n(x)\mlc E(x)\mli G(x)\bigr)\mli \bigl(\ada xF_1(x)\mlc\ldots\mlc\ada xF_n(x)
\mlc\ade xE(x)\mli\ade xG(x)\bigr)$.
\end{lemma}

\begin{proof} Strategy for $n$: Wait till the environment makes a move $c$ in the $\ade xE^*(x)$ component.
 Then make the same move $c$ in $\ade xG^*(x)$, $\ada x\bigl(F_{1}^{*}(x)\mlc\ldots\mlc F_{n}^{*}(x)\mlc E^*(x)\mli G^*(x)\bigr)$ and each 
of the $\ada xF_{i}^{*}(x)$ components. After this series of moves the game will have evolved to 
$\bigl(F_{1}^{*}(c)\mlc\ldots\mlc F_{n}^{*}(c)\mlc E^*(c)\mli G^*(c)\bigr)\mli \bigl(F_{1}^{*}(c)\mlc\ldots\mlc F_{n}^{*}(c)\mlc E^*(c)\mli G^*(c)\bigr)$. Now switch to $\cal CCS$.\vspace{-4pt}\end{proof}

\begin{lemma}\label{oct5b} 
Assume $t$ is free for $x$ in $F(x)$. Then $\uvalid F(t)\mli \ade xF(x)$. Furthermore, there is an effective function that takes any $t$ and constructs an EPM $\cal E$ such that, for any $F(x)$, whenever $t$ is free for $x$ in $F(x)$, \ 
${\cal E}\uvalid F(t)\mli \ade xF(x)$.
\end{lemma}

\begin{proof} Strategy: Let $c=e(t)$. Read $c$ from the valuation tape if necessary. Then make the move `$2.c$', bringing the game down to $F^*(c)\mli F^*(c)$. Then switch to $\cal CCS$.  \end{proof}

\begin{lemma}\label{oct5c} 
Assume $F$ does not contain $x$. Then
$\uvalid F\mli \ada xF$. Furthermore, there is an EPM $\cal E$ such that, for any 
$F$ and $x$, as long as $F$ does not contain $x$, \ ${\cal E}\uvalid F\mli \ada xF$. 
\end{lemma}

\begin{proof} In this case we prefer to explicitly write the usually suppressed parameter $e$. Consider an arbitrary $F$ not containing $x$, and an arbitrary interpretation $^*$ admissible for $F\mli \ada xF$. The formula $F\mli \ada xF$ contains $x$ yet $F$ does not. From the definition of admissibility and with a little thought we can see that  $F^*$ does not depend on $x$. In turn, this means --- as can be seen with some thought --- that the move $c$ by the environment 
(whatever constant $c$) in $e[\ada x F^*]$ brings this game down to 
$e[F^*]$. With this observation in mind, the following strategy can be seen to be successful:
Wait till the environment makes the move `$2.c$' for some constant $c$. Then read 
the sequence `$1.\alpha_1$', \ldots, `$1.\alpha_n$' of (legal) moves possibly made by the environment before it made the move `$2.c$', and make the $n$ moves `$2.\alpha_1$', \ldots, `$2.\alpha_n$'. It can be seen that now the original 
game $e[F^*]\mli e[\ada xF^*]$ will have been brought down to $\seq{\Phi}e[F^*]\mli\seq{\Phi}e[F^*]$, where 
$\Phi=\seq{\pp \alpha_1,\ldots,\pp\alpha_n}$. So, switching to $\cal CCS$ at this point guarantees success.
\end{proof}

\begin{lemma}\label{oct99} 
Assume $F(x)$ does not contain $y$. Then $\uvalid \ada yF(y)\mli\ada xF(x)$ and $\uvalid \ade xF(x)\mli \ade yF(y)$. 
In fact, ${\cal CCS}\uvalid \ada yF(y)\mli\ada xF(x)$ and ${\cal CCS}\uvalid \ade xF(x)\mli \ade yF(y)$.
\end{lemma}

\begin{proof} Assuming that $F(x)$ does not contain $y$ and analyzing the relevant definitions, it is not hard to see that, for any interpretation $^*$ admissible for $\ada yF(y)\mli\ada xF(x)$ and/or $\ade xF(x)\mli \ade yF(y)$,  we simply have $\bigl(\ada yF(y)\bigr)^*=\bigl(\ada x F(x)\bigr)^*$ and $\bigl(\ade yF(y)\bigr)^*=\bigl(\ade x F(x)\bigr)^*$. So, in both cases we deal with a game of the form $A\mli A$, for which the ordinary copy-cat strategy is successful.
\end{proof}

Our proof of the following lemma is fairly long, for which reason it is given separately in Section \ref{prf}:
\begin{lemma}\label{l5}
$\uvalid \st F\mli \st\st F$. Furthermore, there is an EPM $\cal E$ such that, for any $F$, \ ${\cal E}\uvalid \st F\mli \st\st F$.
\end{lemma}

\section{Proof of Lemma \ref{l5}\label{prf}}

Roughly speaking,  the uniform solution $\cal E$ for $\st F\mli\st\st F$ that we are going to construct essentially uses a copy-cat strategy between the antecedent and the consequent. However, this strategy cannot be applied directly in the form of our kind old friend $\cal CCS$. The problem is that the underlying tree-structure (see Remark \ref{newrem}) of the multiset of (legal) runs of $F^*$ that is being generated in the antecedent should be a simple tree $T$,
while in the consequent it is in fact what can be called a tree $T''$ of trees. 
The trick that $\cal E$ uses is that it sees each edge of $T$ in one of two --- blue or yellow --- colors. This allows 
$\cal E$ to associate with each branch $y$ of a branch $x$ of $T''$ a single branch $z$ of $T$, and vice versa. Specifically, with $x,y,z$ being (encoded by) bit strings, 
$z$ is such that the subsequence of its blue-colored bits (=edges) coincides with $x$, and the subsequence of its yellow-colored bits coincides with $y$. By appropriately ``translating" and ``copying" in the antecedent the replicative moves made by the 
environment in the consequent, such an isomorphism between the branches of $T$ and the branches of branches of $T''$ can be successfully maintained throughout the play. With this 
one-to-one correspondence in mind, every time the environment makes a (nonreplicative) move in the position(s) of $F^*$  represented by a leaf or a set of leaves of $T$, the machine repeats the same move in the position(s) represented by the corresponding leaf-of-leaf or leaves-of-leaves of $T''$, and vice versa. The effect achieved by this strategy is that 
the multisets of all runs of $F^*$ in the antecedent and in the consequent of $\st F^*\mli\st\st F^*$ are identical, which,
of course, guarantees a win for $\cal E$.

Let us now define things more precisely. A {\bf colored bit} $b$ is a pair $(c,d)$, where $c$, called the {\bf content} of $b$, is in $\{0,1\}$, and $d$, called the {\bf color} of $b$, is in \{{\em blue,yellow}\}. We will be using the notation $\blu{c}$ (``blue $c$") for the colored bit whose content is $c$ and color is {\em blue}, and $\yel{c}$ (``yellow $c$") for the bit whose content is $c$ and color is {\em yellow}. 

A {\bf colored bit string} is a finite or infinite sequence of colored bits. 
Consider a colored bit string $v$. The {\bf content} of $v$ is the result of ``ignoring the colors" in $v$, i.e. 
replacing every bit of $v$ by the content of that bit. The {\bf blue content} of $v$ is the content of the string that results from deleting in $v$ all but blue bits. {\bf Yellow content} is defined similarly. 
We use $\cont{v}$, $\blu{v}$ and $\yel{v}$ to denote the content, blue content and yellow content of $v$, respectively. 
Example: if $v=\blu{1}\yel{0}\yel{0}\blu{0}\yel{1}$, we have $\cont{v}=10001$, $\blu{v}=10$ and $\yel{v}=001$. 
As in the case of ordinary bit strings, $\epsilon$ stands for the empty colored bit string. And, for colored bit strings $w$ and $u$, $w\preceq u$ again means that $w$ is a (not necessarily proper) initial segment of $u$.

\begin{definition}\label{coltree}
A 
{\bf colored tree} is a set $T$ of colored bit strings, called its {\bf branches}, such that the following conditions are satisfied:

a) The set $\{\cont{v}\ |\ v\in T\}$ --- that we denote by $\cont{T}$ --- is a tree in the sense of Convention \ref{tree}. 

b) For any $w,u\in T$, if $\cont{w}=\cont{u}$, then $w=u$.

c) For no $v\in T$ do we have $\{v\blu{0},v\yel{1}\}\subseteq T$ or $\{v\yel{0},v\blu{1}\}\subseteq T$.

A branch $v$ of $T$ is said to be a {\bf leaf} iff $\cont{v}$ is a leaf of $\cont{T}$.
\end{definition}

 When represented in the style of Figure 1 of \cite{Jap03} (page 36), a colored tree will look like an ordinary tree, with the only difference that now every edge will have one of the colors {\em blue} or {\em yellow}. Also, by condition (c), both of the outgoing 
edges (``sibling" edges) of any non-leaf node will have the same color. 
  
\begin{lemma}\label{sep19}
Assume $T$ is a colored tree, and $w,u$ are branches of $T$ with $\blu{w}\preceq \blu{u}$ and $\yel{w}\preceq \yel{u}$. Then 
$w\preceq u$.
\end{lemma}

\begin{proof} Assume $T$ is a colored tree, $w,u\in T$, and $w\not\preceq u$. We want to show that then $\blu{w}\not\preceq \blu{u}$ or $\yel{w}\not\preceq\yel{u}$. Let $v$ be the longest common initial segment of $w$ and $u$, so we have $w=vw'$ and $u=vu'$ for some 
$w',u'$ such that $w'$ is nonempty and $w'$ and $u'$ do not have a nonempty common initial segment. 
Assume the first bit of $w'$ is $\blu{0}$ (the cases when it is $\blu{1}$, $\yel{0}$ or $\yel{1}$, of course, will be similar). If $u'$ is empty, then $w$ obviously contains more blue bits than $u$ does, and we are done. Assume now $u'$ is nonempty, in particular, $b$ is the first bit of $u'$. Since $w'$ and $u'$ do not have a nonempty common initial segment, $b$ should be different from $\blu{0}$. By condition (b) of Definition \ref{coltree}, the content of $b$ cannot be $0$ (for otherwise we would have $v\blu{0}=vb$ and hence $b=\blu{0}$). Consequently, $b$ is either $\blu{1}$ or $\yel{1}$. The case $b=\yel{1}$ is ruled out by condition (c) of Definition \ref{coltree}. Thus, $b=\blu{1}$. But the blue content of $v\blu{0}$ is $\blu{v}0$ while the blue content of $v\blu{1}$ is $\blu{v}1$. Taking into account the obvious fact that 
the former is an initial segment of $\blu{w}$ and the latter is an initial segment of $\blu{u}$, we find $\blu{w}\not\preceq \blu{u}$.
\end{proof}

Now comes a description of our EPM $\cal E$. At the beginning, this machine creates a record $T$ of the type `finite colored tree', and initializes it to $\{\epsilon\}$. After that, $\cal E$ follows the following procedure:\vspace{7pt}

{\bf Procedure} LOOP: Keep granting permission until the adversary makes a move $\gamma$. If $\gamma$ satisfies the conditions of one of the following four cases, act as the corresponding case prescribes. Otherwise go to an infinite loop in a permission state. 

{\em Case (i):}   $\gamma=2.\col{w}$ for some bit string $w$. Let $v_1,\ldots,v_k$ be\footnote{In each of the four cases 
we assume that the list $v_1,\ldots,v_k$ is arranged lexicographically.} all of the leaves $v$ of $T$ with $w=\blu{v}$. Then make the moves 
$1.\col{\cont{v}_1},\ldots,1.\col{\cont{v}_k}$, update $T$ to $T\cup\{v_1\blu{0},v_1\blu{1},\ldots,v_k\blu{0},v_k\blu{1}\}$, and repeat LOOP.

{\em Case (ii):} $\gamma=2.w.\col{u}$ for some bit strings $w,u$. Let $v_1,\ldots,v_k$ be all of the leaves $v$ of $T$ 
such that $w\preceq \blu{v}$ 
and $u=\yel{v}$. Then make the moves 
$1.\col{\cont{v}_1},\ldots,1.\col{\cont{v}_k}$, update $T$ to $T\cup\{v_1\yel{0},v_1\yel{1},\ldots,v_k\yel{0},v_k\yel{1} \}$, and repeat LOOP.

{\em Case (iii):} $\gamma=2.w.u.\alpha$ for some bits strings $w,u$ and move $\alpha$. Let $v_1,\ldots,v_k$ be all of the leaves $v$ of $T$ 
such that 
$w\preceq \blu{v}$ and $u\preceq\yel{v}$. Then make the moves $1.\cont{v}_1.\alpha,\ldots,1.\cont{v}_k.\alpha$, and repeat LOOP.

{\em Case (iv):} $\gamma=1.w.\alpha$ for some bit string $w$ and move $\alpha$. 
Let $v_1,\ldots,v_k$ be all of the leaves $v$ of $T$ 
with $w\preceq \cont{v}$.  
Then make the moves 
$2.\blu{v}_1.\yel{v}_1.\alpha,\ldots,2.\blu{v}_k.\yel{v}_k.\alpha$, and repeat LOOP.

\

Fix any interpretation $^*$, valuation 
$e$ and $e$-computation branch $B$ of $\cal E$. Let $\Theta$ be the run spelled by $B$. From the above description it is immediately clear that $B$ is a fair. Hence, in order to show that $\cal E$ wins, 
it would be sufficient to show that $\win{\sti F^*\mli \sti\sti F^*}{e}\seq{\Theta}=\pp$.  Notice that the work of $\cal E$  
does not depend on $e$. And, as $e$ is fixed, we can safely and unambiguously omit this parameter (as we often did in the previous section) in expressions
such as  $e[A]$, $\legal{A}{e}$ or $\win{A}{e}$ and just write or say $A$, $\legal{A}{}$ or $\win{A}{}$. 
Of course, $\cal E$ is interpretation-blind, so as long as it wins $\st F^*\mli \st\st F^*$, it is a uniform solution for 
$\st F\mli \st\st F$.

Let $N=\{1,\ldots,m\}$ if LOOP is iterated the finite number $m$ of times in $B$, and $N=\{1,2,3,\ldots\}$ otherwise.  For $i\in N$, we 
let $T_i$ denote the value of record $T$ at the beginning of the $i$th iteration of LOOP; $\Theta_i$ will mean the 
initial segment of $\Theta$ consisting of the moves made by the beginning of the $i$th iteration of LOOP.
 Finally, $\Phi_i$ will stand for $\rneg\Theta_{i}^{1.}$ and $\Psi_i$ for $\Theta_{i}^{2.}$.

From the description of LOOP it is immediately obvious that, 
for each $i\in N$, $T_i$ is a finite colored tree, and that $T_1\subseteq T_2\subseteq \ldots\subseteq T_i$. In our subsequent reasoning we will implicitly rely on this fact.

\begin{lemma}\label{sep21b}
For every $i$ with $i\in N$, we have:\vspace{3pt}

a) $\Phi_i$ is prelegal and $\tree{\sti F^*}{\Phi_i}=\cont{T_i}$.

b) $\Psi_i$ is prelegal.

c) For every leaf $x$ of $\tree{\sti\sti F^*}{\Psi_i}$, $\Psi_{i}^{\preceq x}$ is prelegal.

d) For every leaf $z$ of $T_i$, $\blu{z}$ is a leaf of $\tree{\sti \sti F^*}{\Psi_{i}}$ and $\yel{z}$ is a leaf of 
$\tree{\sti F^*}{\Psi_{i}^{\preceq \blu{z}}}$.

e) For every leaf $x$ of $\tree{\sti \sti F^*}{\Psi_{i}}$ and every leaf $y$ of $\tree{\sti F^*}{\Psi_{i}^{\preceq x}}$, 
there is a leaf $z$ of $T_i$ such that $x=\blu{z}$ and $y=\yel{z}$.  By Lemma \ref{sep19}, such a $z$ is unique.

f) For every leaf $z$ of $T_i$, $\Phi_{i}^{\preceq \cont{z}}=(\Psi_{i}^{\preceq \blu{z}})^{\preceq \yel{z}}$.

g) $\Theta_i$ is a legal position of $\st F^*\mli \st\st F^*$; hence, $\Phi_i\in\legal{\sti F^*}{}$ and 
$\Psi_i\in\legal{\sti\sti F^*}{}$.
\end{lemma}

\begin{proof} We proceed by induction on $i$. The basis case with $i=1$ is rather straightforward for each clause of the lemma and we do not discuss it. For the induction step, assume $i+1\in N$, and the seven clauses of the lemma are true for $i$. 

{\em Clause (a):} By the induction hypothesis, $\Phi_i$ is prelegal and $\tree{\sti F^*}{\Phi_i}=\cont{T}_i$. 
Assume first that the $i$th iteration of LOOP deals with Case (i), so that $\Phi_{i+1}=\seq{\Phi_i,\oo\col{\cont{v}_1},\ldots,\oo\col{\cont{v}_k}}$. Each of $\cont{v}_1,\ldots,\cont{v}_k$ is a leaf of
$\cont{T}_i$, i.e. a leaf of $\tree{\sti F^*}{\Phi_i}$. This guarantees that $\Phi_{i+1}$ is prelegal.  Also,  
by the definition of function {\em Tree}, we have   
$\tree{\sti F^*}{\Phi_{i+1}}=\tree{\sti F^*}{\Phi_i}\cup\{\cont{v}_10,\cont{v}_11,\ldots,\cont{v}_k0,\cont{v}_k1\}$. But the latter is nothing but $\cont{T}_{i+1}$ as can be seen from the description of how Case (i) updates $T_i$ to $T_{i+1}$. 
A similar argument applies when the $i$th iteration of LOOP deals with Case (ii). Assume now the $i$th iteration of LOOP 
deals with Case (iii). Note that the moves made in the antecedent of $\st F^*\mli\st\st F^*$ 
(the moves that bring $\Phi_i$ to $\Phi_{i+1}$) are nonreplicative --- specifically, look like 
$\cont{v}.\alpha$ where $\cont{v}\in \cont{T}_i=\tree{}{\Phi_i}$. Such moves yield prelegal positions and do not change 
the value of {\em Tree}, so that  $\tree{\sti F^*}{\Phi_i}=\tree{\sti F^*}{\Phi_{i+1}}$. It remains to note that 
$T$ is not updated in this subcase, so that we also have $\cont{T}_{i+1}=\cont{T}_{i}$. Hence $\tree{}{\Phi_{i+1}}=\cont{T}_{i+1}$. Finally, suppose the $i$th iteration of LOOP 
deals with Case (iv). It is the environment who moves in the 
antecedent of $\st F^*\mli\st\st F^*$, and does so before the machine makes any moves. Then the clean environment assumption 
--- in conjunction with the induction hypothesis --- implies that such a move cannot bring $\Phi_i$ to an illegal position of $\st F^*$ and hence cannot bring it to a non-prelegal position. So, $\Phi_{i+1}$ is prelegal. As for 
$\tree{}{\Phi_{i+1}}=\cont{T}_{i+1}$, it holds for the same reason as in the previous case. 

{\em Clause (b):} If the $i$th iteration of LOOP deals with Case (i), (ii) or (iii), it is the environment who moves in the 
consequent of $\st F^*\mli\st\st F^*$, and the clean environment assumption guarantees that $\Psi_{i+1}$ is prelegal. Assume now that the $i$th iteration of LOOP deals with Case (iv), so that  
$\Psi_{i+1}=\seq{\Psi_i,\pp\blu{v}_1.\yel{v}_1.\alpha,\ldots,\pp\blu{v}_k.\yel{v}_k.\alpha}$.
By the induction hypothesis for clause (d), 
each $\blu{v}_j$ ($1\leq j\leq k$) is a leaf of $\tree{\sti\sti F^*}{\Psi_i}$, so adding the moves $\pp\blu{v}_1.\yel{v}_1.\alpha,\ldots,\pp\blu{v}_k.\yel{v}_k$ does not 
bring $\Psi_i$ to a non-prelegal position, nor does it modify $\tree{\sti\sti F^*}{\Psi_i}$ because the moves
are nonreplicative. Hence $\Psi_{i+1}$ is prelegal.

{\em Clause (c):} Just as in the previous clause, when the $i$th iteration of LOOP deals with Case (i), (ii) or (iii), the 
desired conclusion follows from the clean environment assumption.  Assume now that the $i$th iteration of LOOP deals with Case (iv). Consider any leaf $x$ of $\tree{\sti\sti F^*}{\Psi_{i+1}}$. As noted when discussing Case (iv) in the proof of Clause (b), $\tree{\sti\sti F^*}{\Psi_i}=\tree{\sti\sti F^*}{\Psi_{i+1}}$, so $x$ is also a leaf of 
$\tree{\sti\sti F^*}{\Psi_i}$. Therefore, 
if $\Psi_{i+1}^{\preceq x}=\Psi_{i}^{\preceq x}$, the conclusion that $\Psi_{i+1}^{\preceq x}$ is  prelegal 
follows from the induction hypothesis. Suppose now $\Psi_{i+1}^{\preceq x}\not=\Psi_{i}^{\preceq x}$. Note that then  $\Psi_{i+1}^{\preceq x}$ looks like $\seq{\Psi_{i}^{\preceq x},\pp y_1.\alpha,\ldots,\pp y_m.\alpha}$, where for each $y_j$ ($1\leq j\leq m$)  
we have $\blu{z}=x$ and $\yel{z}=y_j$ for some leaf $z$ of $T_i$. By the induction hypothesis for clause (d), 
each such $y_j$ is a leaf of $\tree{\sti\sti F^*}{\Psi_{i}^{\preceq x}}$. By the induction hypothesis for the present clause, 
$\Psi_{i}^{\preceq x}$ is prelegal. Adding to such a position the nonreplicative moves 
$\pp y_1.\alpha,\ldots,\pp y_m.\alpha$ --- where the $y_j$ are leaves of $\tree{}{\Psi_{i}^{\preceq x}}$ --- cannot bring it to a non-prelegal position. Thus, $\Psi_{i+1}^{\preceq x}$ remains prelegal. 

{\em Clauses (d) and (e):} If the $i$th iteration of LOOP deals with Cases (iii) or (iv), $T_i$ is not modified, and no moves of the form $\col{x}$ or $x.\col{y}$ (where $x,y$ are bit strings) are made in the consequent of 
$\st F^*\mli\st\st F^*$, so $\tree{\sti \sti F^*}{\Psi_i}$ and $\tree{ \sti F^*}{\Psi_{i}^{\preceq x}}$ (any leaf $x$ of
$\tree{}{\Psi_i}$) are not affected, either. Hence Clauses (d) and (e) for $i+1$ are automatically inherited from the induction hypothesis for these clauses. 
This inheritance also takes place --- even if no longer ``automatically" --- when the $i$th iteration of LOOP deals with Case (i) or (ii). This can be verified by a routine analysis of how Cases (i) and (ii) modify $T_i$ and 
the other relevant parameters. Details are left to the reader.  

{\em Clause (f):} Consider any leaf $z$ of $T_{i+1}$. When the $i$th iteration of LOOP deals with Case (i) or (ii), no moves of the form $x.\alpha$ 
 are made in the antecedent of $\st F^*\mli\st\st F^*$, and no moves of the form $x.y.\alpha$ are made in the consequent (any bit strings $x,y$). Based on this, it is easy to see that for all
bit strings $x,y$ --- including the case $x=\blu{z}$ and $y=\yel{z}$ --- we have $\Phi_{i+1}^{\preceq x}=\Phi_{i}^{\preceq x}$ and $(\Psi_{i+1}^{\preceq x})^{\preceq y}=
(\Psi_{i}^{\preceq x})^{\preceq y}$. Hence clause (f) for $i+1$ is inherited from the same clause for $i$. Now suppose  
the $i$th iteration of LOOP deals with Case (iii). Then $T_{i+1}=T_i$ and hence $z$ is also a leaf of $T_i$. From the description of Case (iii) one can easily see that if $w\not\preceq\blu{z}$ or $u\not\preceq\yel{z}$, we have
$\Phi_{i+1}^{\preceq \cont{z}}=\Phi_{i}^{\preceq \cont{z}}$ and 
$(\Psi_{i+1}^{\preceq\blu{z}})^{\preceq \yel{z}}=(\Psi_{i}^{\preceq\blu{z}})^{\preceq \yel{z}}$, 
so the equation $\Phi_{i+1}^{\preceq \cont{z}}=(\Psi_{i+1}^{\preceq\blu{z}})^{\preceq \yel{z}}$ is true by the induction hypothesis;
and if $w\preceq\blu{z}$ and $u\preceq\yel{z}$, then  $\Phi_{i+1}^{\preceq \cont{z}}=\seq{\Phi_{i}^{\preceq\cont{z}},\oo\alpha}$ and $(\Psi_{i+1}^{\preceq\blu{z}})^{\preceq\yel{z}}=\seq{(\Psi_{i}^{\preceq\blu{z}})^{\preceq\yel{z}},\oo\alpha}$. But, by the induction hypothesis, $\Phi_{i}^{\preceq\cont{z}}=(\Psi_{i}^{\preceq\blu{z}})^{\preceq\yel{z}}$. Hence 
$\Phi_{i+1}^{\preceq \cont{z}}=(\Psi_{i+1}^{\preceq\blu{z}})^{\preceq\yel{z}}$.
A similar argument applies when the $i$th iteration of LOOP deals with Case (iv).

{\em Clause (g):} 
Note that all of the moves made in any of Cases (i)-(iv) of LOOP have the prefix `$1.$' or `$2.$', i.e. are made either in the antecedent or the consequent of $\st F^*\mli\st\st F^*$. Hence, in order to show that $\Theta_{i+1}$ is a legal position of 
 $\st F^*\mli \st\st F^*$, it would suffice to verify that $\Phi_{i+1}\in\legal{\sti F^*}{}$ and 
$\Psi_{i+1}\in\legal{\sti\sti F^*}{}$. 

Suppose the $i$th iteration of LOOP deals with Case (i) or (ii). The clean environment assumption guarantees that  $\Psi_{i+1}\in\legal{\sti\sti F^*}{}$. In the antecedent of $\st F^*\mli\st\st F^*$ only replicative moves are made. 
Replicative moves can yield an illegal position ($\Phi_{i+1}$ in our case) of a $\st$-game only if they yield a non-prelegal 
position.  But, by clause (a), $\Phi_{i+1}$ is prelegal. Hence it is a legal position of $\st F^*$. 

Suppose now the $i$th iteration of LOOP deals with Case (iii). Again, that  
$\Psi_{i+1}\in\legal{\sti\sti F^*}{}$ is guaranteed by the clean environment assumption. So, we only need to verify that $\Phi_{i+1}\in\legal{\sti F^*}{}$. By clause (a), this position is prelegal. So, it remains to see that, for any leaf 
$y$ of $\tree{\sti F^*}{\Phi_{i+1}}$, $\Phi_{i+1}^{\preceq z}\in\legal{\sti F^*}{}$.  
Pick an arbitrary leaf $y$ of $\tree{\sti F^*}{\Phi_{i+1}}$  --- i.e., by clause (a), of $\cont{T}_{i+1}$. Let $z$ be the leaf of $T_{i+1}$ with $y=\cont{z}$. We already know that $\Psi_{i+1}\in\legal{\sti\sti F^*}{}$. By clause (d), we also know that $\blu{z}$ is a leaf of $\tree{\sti \sti F^*}{\Psi_{i+1}}$. Consequently, 
$\Psi_{i+1}^{\preceq \blu{z}}\in\legal{\sti F^*}{}$. Again by clause (d), $\yel{z}$ is a leaf of 
$\tree{\sti F^*}{\Psi_{i+1}^{\preceq \blu{z}}}$. Hence, $(\Psi_{i+1}^{\preceq \blu{z}})^{\preceq \yel{z}}$ should be a legal position of $F^*$.
But, by clause (f), $\Phi_{i+1}^{\preceq \cont{z}}=(\Psi_{i+1}^{\preceq \blu{z}})^{\preceq \yel{z}}$. Thus,
$\Phi_{i+1}^{\preceq y}\in\legal{F^*}{}$.

Finally, suppose the $i$th iteration of LOOP deals with Case (iv). By the clean environment assumption, $\Phi_{i+1}\in\legal{\sti F^*}{}$. Now consider $\Psi_{i+1}$. This position 
is  prelegal by clause (b). 
So, in order for $\Psi_{i+1}$  to be a legal position 
of $\st \st F^*$, for every leaf $x$ of 
$\tree{\sti\sti F^*}{\Psi_{i+1}}$ we should have $\Psi_{i+1}^{\preceq x}\in\legal{\sti F^*}{}$. Consider an arbitrary such leaf $x$. By clause (c), $\Psi_{i+1}^{\preceq x}$ is prelegal. Hence, a sufficient condition for 
$\Psi_{i+1}^{\preceq x}\in\legal{\sti F^*}{}$ is that, for every leaf $y$ of $\tree{}{\Psi_{i+1}^{\preceq x}}$,  $(\Psi_{i+1}^{\preceq x})^{\preceq y}\in\legal{F^*}{}$. So, let $y$ be an arbitrary such leaf.
 By clause (e), there is a leaf $z$ of $T_{i+1}$ such that $\blu{z}=x$ and $\yel{z}=y$.
Therefore, by clause (f), $\Phi_{i+1}^{\preceq \cont{z}}=(\Psi_{i+1}^{\preceq x})^{\preceq y}$. But we know that $\Phi_{i+1}\in\legal{\sti F^*}{}$ and hence (with clause (a) in mind) $\Phi_{i+1}^{\preceq \cont{z}}\in\legal{F^*}{}$. Consequently,
$(\Psi_{i+1}^{\preceq x})^{\preceq y}\in\legal{F^*}{}$.
\end{proof}

\begin{lemma}\label{sep21c} 
For every finite initial segment $\Upsilon$ of $\Theta$, there is $i\in N$ such that $\Upsilon$ is a $($not necessarily proper$)$ initial segment of 
$\Theta_i$ and hence of every $\Theta_j$ with $i\leq j\in N$. 
\end{lemma}

\begin{proof} 
The statement of the lemma is straightforward when there are infinitely many iterations of LOOP, for each iteration 
adds a nonzero number of new moves to the run and hence there are arbitrarily long $\Theta_i$s, each of them  being an initial segment of $\Theta$. Suppose now 
 LOOP is iterated a finite number $m$ of times. It would be (necessary and) sufficient to verify that in this case $\Theta=\Theta_m$, i.e. no moves are made during the last iteration of LOOP. But this is indeed so. From the description of LOOP we see that the machine does not make any moves during a given iteration unless the environment makes a move $\gamma$ first. So, assume $\oo$ makes move $\gamma$ during the $m$th iteration of LOOP. By the clean environment assumption, 
we should have $\seq{\Theta_m,\oo\gamma}\in\legal{\sti F^*\mli\sti\sti F^*}{}$. It is 
easy to see that such a $\gamma$ would have to satisfy the conditions of one of the Cases (i)-(iv) of LOOP. But then there would be an $(m+1)$th iteration of LOOP, contradicting out assumption that there are only $m$ iterations. \end{proof}

Let us use $\Phi$ and $\Psi$ to denote $\rneg\Theta^{1.}$ and $\Theta^{2.}$, respectively. Of course, the statement of 
Lemma \ref{sep21c} is true for $\Phi$ and $\Psi$ (instead of $\Theta$) as well.
Taking into account that, by definition, a given run is legal if all of its finite initial segments are legal, the following fact is an immediate corollary of Lemmas \ref{sep21c} and \ref{sep21b}(g):

\begin{equation}\label{sep21d}
\mbox{\em $\Theta\in\legal{\sti F^*\mli \sti\sti F^*}{}$. Hence, $\Phi\in\legal{\sti F^*}{}$ and $\Psi\in\legal{\sti\sti F^*}{}$.}
\end{equation}

To complete our proof of Lemma \ref{l5}, we need to show that  $\win{\sti F^*\mli \sti\sti F^*}{}\seq{\Theta}=\pp$.
With (\ref{sep21d}) in mind, if $\win{\sti\sti F^*}{}\seq{\Psi}=\pp$, we are done. Assume now  
$\win{\sti\sti F^*}{}\seq{\Psi}=\oo$. Then, by the definition of $\st$, 
there is an infinite bit string $x$ such that   
$\Psi^{\preceq x}$ is a legal but lost (by $\pp$) run of $\st F^*$. This means that, for some infinite bit string $y$, 
\begin{equation}\label{sp19}
\win{F^*}{}\seq{(\Psi^{\preceq x})^{\preceq y}}=\oo.
\end{equation}
 Fix these $x$ and $y$. For each $i\in N$, let $x_i$ denote the (obviously unique) leaf of $\tree{\sti \sti F^*}{\Psi_{i}}$ such that $x_i\preceq x$; and let $y_i$ denote the (again unique) leaf of $\tree{\sti F^*}{\Psi_{i}^{\preceq x_i}}$ such that $y_i\preceq y$. 
Next, let 
$z_i$ denote the leaf of $T_i$ with $\blu{z}_i=x_i$ and $\yel{z}_i=y_i$. According to Lemma \ref{sep21b}(e), 
such a $z_i$ exists and is unique. 

Consider any $i$ with $i+1\in N$. Clearly $x_i\preceq x_{i+1}$ and $y_i\preceq y_{i+1}$. By our choice of the $z_j$,
we then have $\blu{z}_i\preceq\blu{z}_{i+1}$ and $\yel{z}_i\preceq\yel{z}_{i+1}$. Hence, by Lemma \ref{sep19}, 
$z_i\preceq z_{i+1}$. Let us fix an infinite bit string $z$ such that for every $i\in N$, $\cont{z}_i\preceq z$. Based on the 
just-made observation that we always have  $z_i\preceq z_{i+1}$,  such a $z$ exists.

In view of Lemma \ref{sep21c}, Lemma \ref{sep21b}(f) easily allows us to find that $\Phi^{\preceq z}=(\Psi^{\preceq x})^{\preceq y}$. Therefore, by (\ref{sp19}), $\win{F^*}{}\seq{\Phi^{\preceq z}}=\oo$. By the definition of $\st$, this means that $\win{\sti F^*}{}\seq{\Phi}=\oo$. Hence, by the definition of $\mli$ and with (\ref{sep21d}) in mind, 
$\win{\sti F^*\mli\sti\sti F^*}{}\seq{\Theta}=\pp$. Done.

\section{Proof of Theorem \ref{main}}\label{smain}

Now we are ready to prove our main Theorem \ref{main}. Consider an arbitrary sequent $S$ with $\hint\vdash S$. 
By induction on the $\hint$-derivation of $S$, we are going to show that $S$ has a uniform solution $\cal E$. 
This is sufficient to conclude that $\hint$ is `uniformly sound'. The theorem also claims `constructive soundness', i.e. that such an $\cal E$ can be effectively built from a given $\hint$-derivation of $S$. This claim of the theorem will be automatically taken care of by the fact that our proof of the existence of $\cal E$ is constructive: the uniform-validity and closure lemmas on which
we rely provide a way for actually constructing a corresponding uniform solution. With this remark in mind and for 
the considerations of readability, in what follows we only talk about uniform validity without explicitly mentioning 
uniform solutions for the corresponding formulas/sequents and without explicitly showing how to construct such solutions. Also, we no longer use $\Rightarrow$ or $\intimpl$, seeing each sequent $\s{F}\Rightarrow K$ as the formula $\s{\st F}\mli K$ and each subformula $E_1\intimpl E_2$ of such a formula as $\st E_1\mli E_2$. This is perfectly legitimate because, by definition, $(\s{F}\Rightarrow K)^*=(\s{\st F}\mli K)^*$ and $(E_1\intimpl E_2)^*=(\st E_1\mli E_2)^*$.

 There are 15 cases to consider, corresponding to the 15 possible rules that might have been used at the last step of an $\hint$-derivation of $S$, with $S$ being the conclusion of the rule. In each non-axiom case below, ``induction hypothesis" means the assumption that the premise(s) of the corresponding rule is (are) uniformly valid. The goal in each case is to show that the conclusion of the rule is also uniformly valid. ``Modus ponens" should be understood as Lemma \ref{l1}, and ``transitivity" as Lemma \ref{l1c}.\vspace{10pt} 

{\bf Identity:} Immediately from Lemma \ref{l6a}.\vspace{10pt}

{\bf Domination:} Immediately from Lemma \ref{l6b}.\vspace{10pt}
 
{\bf Exchange:} By the induction hypothesis, $\uvalid \s{\st G}\mlc\st E\mlc\st F\mlc\s{\st H}\mli K$. And, by Lemma \ref{l8}(a),  
\(\uvalid (\s{\st G}\mlc\st E\mlc\st F\mlc\s{\st H}\mli K)\mli(\s{\st G}\mlc\st F\mlc\st E\mlc\s{\st H}\mli K).\) 
 Applying  modus ponens yields $\uvalid \s{\st G}\mlc\st F\mlc\st E\mlc\s{\st H}\mli K$.\vspace{10pt}
 
{\bf Weakening:} Similar to the previous case, using Lemma \ref{l8}(b) instead of \ref{l8}(a).\vspace{10pt} 

{\bf Contraction}: By Lemma \ref{l8}(c) (with 
empty $\s{U}$),   
\(\uvalid (\st F\mli\st F\mlc\st F)\mli (\s{\st G}\mlc \st F \mli \s{\st G}\mlc \st F\mlc \st F).\)
And, by Lemma \ref{l6c}, $\uvalid \st F\mli\st F\mlc\st F$. Hence, by modus ponens, 
\(\uvalid \s{\st G}\mlc \st F \mli \s{\st G}\mlc \st F\mlc \st F.\)
But, by the induction hypothesis, $\uvalid\s{\st G}\mlc \st F\mlc \st F\mli K$. Hence, by transitivity, 
$\uvalid \s{\st G}\mlc \st F\mli K$.\vspace{10pt}  

{\bf Right $\intimpl$:} 
 From Lemma \ref{l8}(d),  
$\uvalid \bigl(\s{\st G}\mlc\st F\mli K\bigr)\mli\bigl(\s{\st G}\mli(\st F\mli K)\bigr)$. And, by the induction hypothesis, $\uvalid \s{\st G}\mlc\st F\mli K$. Applying modus ponens, we get  
$\uvalid \s{\st G}\mli(\st F\mli K)$.\vspace{10pt}   

{\bf Left $\intimpl$:} By the induction hypothesis, 
\begin{eqnarray}
& \uvalid  \s{\st G}\mlc \st F\mli K_{1}; &\label{e2}\\ 
& \uvalid  \s{\st H}\mli K_{2}. &  \label{e1}
\end{eqnarray}
Our goal is to show that 
\begin{equation}\label{e3}
\uvalid \s{\st G}\mlc \s{\st H}\mlc \st (\st K_{2}\mli F)\mli K_{1}.
\end{equation}

By Lemma \ref{l10}, (\ref{e1}) implies $\uvalid \st(\s{\st H}\mli K_{2})$. Also, by Lemma 
\ref{l4}, $\uvalid \st(\s{\st H}\mli K_{2})\mli(\st\hspace{1pt}\s{\st H}\mli\st K_{2})$. Applying modus ponens, we get 
$\uvalid \st\hspace{1pt}\s{\st H}\mli\st K_{2}$. Again using Lemma \ref{l10}, we find $\uvalid \st(\st\hspace{1pt}\s{\st H}\mli\st K_{2})$, which, (again)
by Lemma \ref{l4} and modus ponens, implies 
\begin{equation}\label{e4}
\uvalid \st\st\hspace{1pt}\s{\st H}\mli\st\st K_{2}.
\end{equation}

Combining Lemmas \ref{l8}(c) (with empty $\s{W},\s{U}$) and \ref{l5}, by modus ponens, we find $\uvalid \s{\st H}\mli \s{\st \st H}$. Next, by lemma \ref{l4a}, 
$\uvalid \s{\st\st H}\mli\st\vspace{1pt}\s{\st H}$. Hence, by transitivity,  
$\uvalid \s{\st H}\mli \st\vspace{1pt}\s{\st H}$. At the same time, by Lemma \ref{l5}, $\uvalid \st\vspace{1pt}\s{\st H}\mli
\st \st\vspace{1pt}\s{\st H}$. Again by transitivity, $\uvalid \s{\st H}\mli\st \st\vspace{1pt}\s{\st H}$.
 This, together with (\ref{e4}), by transitivity, yields
\begin{equation}\label{e5}
\uvalid \s{\st H}\mli\st\st K_{2}.
\end{equation}

Next, by Lemma \ref{l4}, 
\begin{equation}\label{e9}
\uvalid \st(\st K_{2}\mli F)\mli(\st\st K_{2}\mli\st F).
\end{equation}
From Lemma \ref{l8}(e),  
\[\uvalid \bigl(\st(\st K_{2}\mli F)\mli(\st\st K_{2}\mli\st F)\bigr)\mlc
(\s{\st H}\mli\st\st K_{2})\mli\bigl(\st(\st K_{2}\mli F)\mli(\s{\st H}\mli\st F)\bigr).\] The above, together with (\ref{e9}) and (\ref{e5}), by modus ponens,  yields 
\begin{equation}\label{e6}
\uvalid\st(\st K_{2}\mli F)\mli(\s{\st H}\mli\st F).
\end{equation}
By Lemma \ref{l8}(f), 
\begin{equation}\label{e7}
\uvalid\bigl(\st(\st K_{2}\mli F)\mli(\s{\st H}\mli\st F)\bigr)\mlc
\bigl(\st\s{G}\mlc\st F\mli K_{1}\bigr)\mli
\bigl(\s{\st G}\mlc\s{\st H}\mlc\st(\st K_{2}\mli F)\mli K_{1}\bigr).\end{equation}
From (\ref{e6}), (\ref{e2}) and (\ref{e7}), by modus ponens, we obtain the desired (\ref{e3}).\vspace{10pt}

{\bf  Right $\adc$:} By the induction hypothesis,  $\uvalid \s{\st G}\mli K_1$, \ldots, $\uvalid \s{\st G}\mli K_n$. And, 
from Lemma \ref{l8}(h), 
\(\uvalid (\s{\st G}\mli K_1)\mlc\ldots\mlc (\s{\st G}\mli K_n)\mli(\s{\st G}\mli K_1\adc\ldots\adc K_n).\)
Modus ponens yields 
$\uvalid \s{\st G}\mli K_1\adc\ldots\adc K_n$.\vspace{10pt}

{\bf Left $\adc$:}  By Lemma \ref{l11}(a), $\uvalid \st(F_1\adc\ldots\adc F_n)\mli\st F_i$; and, by Lemma \ref{l8}(c), 
$\uvalid \bigl(\st(F_1\adc\ldots\adc F_n)\mli\st F_i\bigr)\mli
\bigl(\s{\st G}\mlc\st (F_1\adc\ldots\adc F_n)\mli  \s{\st G}\mlc \st F_i\bigr)$.  Modus ponens yields 
$\s{\st G}\mlc\st (F_1\adc\ldots\adc F_n)\mli  \s{\st G}\mlc \st F_i$.
But, by the induction hypothesis, $\uvalid \s{\st G}\mlc \st F_i\mli K$. So, by transitivity, 
$\uvalid \s{\st G}\mlc \st (F_1\adc\ldots\adc F_n)\mli K$.\vspace{10pt}

{\bf Right $\add$:} By the induction hypothesis, $\uvalid \s{\st G}\mli K_i$. According to Lemma \ref{l8}(j), $\uvalid (\s{\st G}\mli K_i)\mli(\s{\st G}\mli K_1\add\ldots\add K_n)$. Therefore, by modus ponens, $\uvalid \s{\st G}\mli K_1\add\ldots\add K_n$.\vspace{10pt}

{\bf Left $\add$:} By the induction hypothesis, $\uvalid\s{ \st G}\mlc\st F_1\mli K$, \ldots, $\uvalid \s{\st G}\mlc\st F_n\mli K$. And, 
by Lemma \ref{l8}(i), 
\(\uvalid (\s{\st G}\mlc \st F_1\mli K)\mlc\ldots\mlc(\s{\st G}\mlc \st F_n\mli K)\mli \bigl(\s{\st G}\mlc (
\st F_1\add\ldots\add \st F_n)\mli K\bigr).\)
Hence, by modus ponens, 
\begin{equation}\label{e10}
\uvalid \s{\st G}\mlc (\st F_1\add\ldots\add \st F_n)\mli K.
\end{equation}  
Next, by Lemma \ref{l8}(c), 
\[\uvalid \bigl(\st(F_1\add\ldots\add F_n)\mli \st F_1\add\ldots\add \st F_n\bigr)
\mli\bigl(\s{\st G}\mlc \st (F_1\add\ldots\add F_n)\mli
\s{\st G}\mlc (\st F_1\add\ldots\add \st F_n)\bigr).\]
But,  by Lemma \ref{l11}(c), 
$\uvalid \st(F_1\add\ldots\add F_n)\mli \st F_1\add\ldots\add \st F_n$. 
Modus ponens yields $\uvalid \s{\st G}\mlc \st (F_1\add\ldots\add F_n)\mli
\s{\st G}\mlc (\st F_1\add\ldots\add \st F_n)$.
From here and (\ref{e10}), by transitivity, $\uvalid \s{\st G}\mlc \st (F_1\add\ldots\add F_n)\mli K$.\vspace{10pt}

{\bf Right $\ada$:} First, consider the case when $\s{\st G}$ is nonempty. By the induction hypothesis, $\uvalid \s{\st G}\mli K(y)$. Therefore, by Lemma \ref{l10a},  
$\uvalid \ada y\bigl(\s{\st G}\mli K(y)\bigr)$ and, by Lemma \ref{oct5a} and modus ponens, 
$\uvalid \ada y\vspace{1pt}\s{\st G}\mli \ada yK(y)$. At the same time, by Lemma \ref{oct5c}, $\uvalid \s{\st G}\mli\ada y\vspace{1pt}\s{\st G}$. By transitivity, we then get $\uvalid \s{\st G}\mli \ada yK(y)$. But, by Lemma \ref{oct99}, $\uvalid \ada yK(y)\mli\ada x K(x)$. Transitivity yields $\uvalid \s{\st G}\mli \ada xK(x)$. The case when $\s{\st G}$ is 
empty is simpler, for then $\uvalid \s{\st G}\mli \ada xK(x)$, i.e. $\uvalid\ada xK(x)$, can be obtained directly from the induction hypothesis by Lemmas \ref{l10a}, \ref{oct99} and modus ponens. \vspace{10pt}

{\bf Left $\ada$:}  
Similar to Left $\adc$, only using Lemma \ref{l11}(b) instead of \ref{l11}(a).\vspace{10pt}

{\bf Right $\ade$:} By the induction hypothesis, $\uvalid \s{\st G}\mli K(t)$. And, by Lemma \ref{oct5b}, 
$\uvalid K(t)\mli \ade xK(x)$. Transitivity yields $\uvalid \s{\st G}\mli \ade xK(x)$.\vspace{10pt}

{\bf Left $\ade$:} By the induction hypothesis, $\uvalid \s{\st G}\mlc\st F(y)\mli K$. This, by Lemma \ref{l10a}, implies 
\(\uvalid \ada y\bigl(\s{\st G}\mlc\st F(y)\mli K\bigr).\) From here, by Lemma \ref{oct5d} and modus ponens, we get
\begin{equation}\label{oct8}
\uvalid \s{\ada y\st G}\mlc\ade y\st F(y)\mli \ade y K.
\end{equation}
By Lemma \ref{l8}(c), 
$\uvalid(\s{\st G \mli\ada y\st G})\mli \bigl(\s{\st G}\mlc\ade y\st F(y)\mli\s{\ada y\st G}\mlc\ade y\st F(y)\bigr)$.
This, together with Lemma \ref{oct5c}, by modus ponens, implies $\s{\st G}\mlc\ade y\st F(y)\mli\s{\ada y\st G}\mlc\ade y\st F(y)$. From here and (\ref{oct8}), by transitivity, 
$\uvalid \s{\st G}\mlc \ade y \st F(y)\mli  K$. But, by Lemmas \ref{oct99}, \ref{l8}(c) and modus ponens, 
$\uvalid \s{\st G}\mlc \ade x \st F(x) \mli \s{\st G}\mlc \ade y \st F(y)$. Hence, by transitivity, 
\begin{equation}\label{oct8a}
\uvalid \s{\st G}\mlc \ade x \st F(x)\mli  K.
\end{equation}
Next, by Lemma \ref{l8}(c), 
\(\uvalid \bigl(\st\ade xF(x)\mli \ade x\st F(x)\bigr)
\mli\bigl(\s{\st G}\mlc \st \ade xF(x)\mli
\s{\st G}\mlc \ade x \st F(x)\bigr).\)
But,  by Lemma \ref{l11}(d), 
$\uvalid \st\ade xF(x)\mli \ade x\st F(x)$. 
Modus ponens yields $\s{\st G}\mlc \st \ade xF(x)\mli
\s{\st G}\mlc \ade x \st F(x)$.
From here and (\ref{oct8a}), by transitivity, $\uvalid \s{\st G}\mlc \st \ade xF(x)\mli K$.

\end{document}